\documentclass[twocolumn,aps,prx,superscriptaddress,floatfix,nofootinbib]{revtex4-2}

\usepackage[T1]{fontenc}
\usepackage{graphicx}
\usepackage{amsmath,amssymb,bm}
\usepackage[dvipsnames]{xcolor}
\usepackage{colortbl,tabularx} 
\usepackage[colorlinks=true, linkcolor=black, citecolor=blue, urlcolor=blue]{hyperref}
\usepackage[capitalize]{cleveref}
\usepackage{float}

 \newcommand{\vpar}{v_\parallel}

 \newcommand{\grad}{\nabla}

\newcommand{\ExB}{\bm E\times\bm B}

\newcommand{\gyacomo}{\textsc{Gyacomo}}
\newcommand{\gkyl}{Gkeyll}

\usepackage{natbib}
\usepackage{graphicx}
\usepackage{bm}
\usepackage{cancel}
\usepackage{float}
\usepackage{algorithm}
\usepackage{algpseudocode}

\begin{document}

\title{High-throughput full-f gyrokinetics of the tokamak boundary}

\author{A.~C.~D. Hoffmann}
\email{ahoffmann@pppl.gov}
\affiliation{Princeton Plasma Physics Laboratory, Princeton, NJ, USA.}
\author{M. Francisquez}
\affiliation{Princeton Plasma Physics Laboratory, Princeton, NJ, USA.}
\author{T.~N. Bernard}
\affiliation{General Atomics, San Diego, CA, USA.}
\author{G.~W. Hammett}
\affiliation{Princeton Plasma Physics Laboratory, Princeton, NJ, USA.}

\date{\today}

\begin{abstract}
    Full-f gyrokinetics of the tokamak boundary have so far been limited to individual, expert-curated cases.
    We report hundreds of concurrent and unsupervised simulations of the edge and scrape-off layer (SOL) in a geometry based on the Tokamak \`a Configuration Variable (TCV), scanning triangularity, elongation and heating power, each evolving turbulence for more than a millisecond, i.e. about two energy confinement times.
    Without tuned transport coefficients or experimental inputs, the ensemble reproduces the range of SOL heat-flux decay lengths measured in TCV, as well as the reduced SOL density fluctuations and the improved confinement of negative triangularity.
    The effect of triangularity on confinement is power dependent.
    At high power, negative triangularity improves the ion energy confinement by weakening the dominant microinstability rather than by a stronger sheared flow.
    At low power, positive triangularity flattens the ion temperature profile across the separatrix, which we attribute to a longer distance between the trapped ions and the cold recycling source.
    Local linear gyrokinetic calculations and the transport coefficients extracted from the turbulent fluxes consistently indicate a trapped-electron-mode regime, in which the conductive heat flux at the separatrix is carried by the electrons and the ion energy transport is convective.
    This article details the simulation setup, the ensemble statistics and the verification of the results, which are also summarized in a companion Letter.
\end{abstract}

\maketitle

\section{Introduction}
\label{sec:introduction}

Magnetic confinement fusion is entering a new phase with public and private efforts aiming to demonstrate net energy gain in the 2030s \citep{Creely2020OverviewTokamak,typeonedesign2025,proximafusiondesign2025,theaenergydesign2025}.
This transition makes the predictive understanding and optimization of turbulent transport more pressing, as turbulence directly sets the achievable energy confinement \citep{wagner1982hmode,Pochelon1999EnergyHeating,Marinoni2019H-modeDIII-D,Adam1976DestabilizationResonances,Liewer1985MeasurementsTransport,Romanelli1989IonTokamaks}.
Despite the progress of quasi-linear and reduced models \citep{Rewoldt1987CollisionalTransport,Staebler2005Gyro-LandauParticles,Bourdelle2007APlasmas,Sun2023AShaping,Avdeeva2023EnergyModels}, an experimentally agnostic prediction of turbulent transport, necessary for developing fusion pilot plants (FPPs), still requires high-fidelity, first-principles numerical simulations.
In this context, gyrokinetic (GK) theory and simulations aim to provide a fundamental model of the nature of turbulence in fusion devices, making highly predictive simulations possible \citep{Catto1978LinearizedGyro-kinetics,Beer1996,Jenko2000,Schekochihin2008GyrokineticSpace,Brizard2007}.

In the early 2000s, nonlinear local, core delta-f GK simulations were accessible to only a handful of people \citep{Dorland1993GyrofluidEffects,Jenko2000}. Now, algorithmic advances and GPU acceleration make these simulations routine, e.g.\ executable on a laptop \citep{Hoffmann2023GyrokineticOperators,Mandell2024GX}, scalable to large parameter scans \citep{Highcock2012Zero-TurbulenceManifold}, and used for self-consistent transport predictions in design studies \citep{Kim2023OptimizationStellarators,guttenfelder2025typeonegkyrokineticpredictions} and optimization loops \citep{Highcock2018gkoptimisation}.
However, such simulations evolve plasma fluctuations on static background profiles in the vicinity of a single magnetic flux surface, requiring closed flux surfaces and scale separation between turbulence and background.
These assumptions break down in the plasma boundary, which encompasses the closed-field-line edge and the open-field-line scrape-off layer (SOL), and is critical to pedestal formation, divertor heat exhaust, fueling, and impurity transport.
The loss of scale separation is particularly acute near marginal stability, where reduced models that do not evolve turbulence self-consistently can break down, both in the core \citep{Gillot2023TheTurbulence} and in the edge \citep{Ashourvan2024}.

Full-f boundary GK simulations lift these restrictions by evolving the entire plasma distribution function across both closed and open field lines, without imposing scale or temporal separation.
Codes such as GENE-X \citep{michels2021}, XGC \citep{chang2017}, ELMFIRE \citep{Korpilo2016elmfire}, GYSELA \citep{grandgirard2016,Caschera2018gyselaIBC,difpradalier2022transportbarrier}, COGENT \citep{dorf2020} and JOREK-GK \citep{becoulet2026jorekgk} enable full-f GK simulations, yet each requires substantial computational resources and expert human intervention, limiting their application to individual cases.
This cost precludes extensive parameter scans, such as the optimization of the plasma shape and of its response to heating power, which are crucial for FPP design.
Plasma shape has been observed to affect transport in TCV \citep{Coda2022EnhancedTCV}, DIII-D \citep{Austin2019AchievementTokamak,Thome_2024} and ASDEX Upgrade \citep{Happel2023OverviewTokamak}, and unconventional shapes, such as negative triangularity \citep{Marinoni2021APlasmas,marinoni2026nondimensionalconfinementscalingsimilar}, are now being considered for FPPs \citep{Rutherford2024MANTA,yuksek2026feasibilitynegativetriangularityequilibria}.
Similarly, the injected power is known to change the transport characteristics of the boundary \citep{Pochelon1999EnergyHeating,Coda2022EnhancedTCV} and to trigger the transition from a low to a high confinement regime \citep{wagner1982hmode,Wagner2007}.
Shaping effects on turbulence have been studied with local GK simulations \citep{Belli2008EffectsTurbulence,Merlo2019TurbulentTriangularity,Duff2022EffectTurbulence,Merlo2023InterplayPlasmas,Balestri2024PhysicalPlasmas,Hoffmann2025InvestigationSimulations}, global core GK simulations \citep{Giannatale2022TriangularitySimulations,Wang2022GlobalPlasmas,DiGiannatale2024SystemSimulations}, fluid boundary simulations \citep{Riva2017PlasmaTurbulence,Laribi2021ImpactSimulations,Lim2023EffectConfigurations} and, in a few cases, full-f boundary GK simulations \citep{Bernard2024PlasmaPlasmas,ulbl2025simulationsedgesolturbulence}, each covering one or a few shapes at a time.

The present work demonstrates that it is now possible to overcome the low-throughput limitation of full-f boundary GK.
Using the open-source \gkyl{} code \citep{gkeyllreadthedoc,Francisquez2026velocity_mappings}, hundreds of full-f boundary GK simulations are performed concurrently and without case-by-case human intervention.
Building upon the predictive framework introduced in \cite{hoffmann2026fullypredictivegyrokineticfullf}, the present simulation campaign studies the effect of plasma shaping and power injection on confinement properties in the boundary of an equilibrium inspired by the Tokamak \`a Configuration Variable (TCV) \citep{Coda2022EnhancedTCV,duval2024tcvteam}.
The shape-power parameter space is scanned with 256 simulations composed of 8 triangularity, 8 elongation, and 4 power injection values.
Each simulation evolves more than 1 millisecond of plasma turbulence, i.e.\ about two energy confinement times of the simulated volume, and the whole ensemble is completed in a single campaign of 5.4 days of wall-clock time and 17,000 GPU node hours on the Perlmutter supercomputer, with restarts handled automatically.
No transport coefficient is manually adjusted, no high-collisionality limit is assumed, as required by fluid models \citep{dudson2009,stegmeir2018,zhu2018gdb,Giacomin2022TheBoundary}, and the setup relies solely on design parameters, i.e.\ the magnetic geometry, the initial plasma profiles, and the energy and particle injection rates \citep{hoffmann2026fullypredictivegyrokineticfullf}.
The \gkyl{} code discretizes the GK equations with a conservative discontinuous Galerkin (DG) scheme \citep{Juno2018DiscontinuousPlasmas}, whose conservation properties and self-consistent numerical dissipation provide the robustness required by unsupervised runs.
Although no experimental measurement enters the setup, the ensemble reproduces the range of the SOL heat-flux decay length \citep{tsui2017heatfluxfeatureTCVsol}, the reduction of SOL density fluctuations at negative triangularity \citep{han2021suppressionfluctuationsNT} and the improved confinement of negative triangularity at high power \citep{Coda2022EnhancedTCV}.
We find that the effect of triangularity on confinement is power dependent, and we identify the dominant instability and the transport channels across the scan.
The time evolution of the distribution functions, moments and electrostatic potential is released as an open simulation ensemble of approximately 75 TB \citep{hoffmann2026tcvmillerscandata}.
A condensed account of the main results is given in a companion Letter \citep{hoffmann2026tcvmillerscan_letter}, and the present article provides the complete description of the setup, the ensemble statistics, and the verification of the results.

The remainder of this paper is organized as follows.
Section~\ref{sec:model} describes the gyrokinetic model and Sec.~\ref{sec:scan_and_database} the parameter scan and the numerical setup.
Section~\ref{sec:results} verifies the quasi-stationarity of the turbulence, gives a qualitative overview of the turbulent boundary, and analyzes the effect of shaping and power injection on the profiles, the radial electric field and the gradients.
Section~\ref{sec:instabilities} compares the dominant microinstability predicted by local linear gyrokinetic simulations with the turbulent transport coefficients of the ensemble.
Section~\ref{sec:trapped_ions} presents a trapped-ion interpretation of the flat SOL ion temperature profile of positive triangularity at low power.
Section~\ref{sec:conclusions} summarizes the findings, discusses the limitations, and outlines future extensions and uses of the simulation ensemble.

\section{Gyrokinetic model}
\label{sec:model}

The simulations in this work use the model and framework described and validated against TCV experiments in \cite{hoffmann2026fullypredictivegyrokineticfullf}, to which we refer for a complete description. We give a brief summary here and provide the detailed form of the collision operator, adaptive sources, and boundary conditions in Appendix~\ref{sec:appendix_setup}.
We use the \gkyl{} \citep{gkeyllreadthedoc} open-source software to solve the full-$f$ GK equations for deuterium ions and electrons in the electrostatic, long-wavelength limit ($k_\perp \rho_i \ll 1$) \citep{Francisquez2026velocity_mappings}, i.e.
\begin{align}
    \frac{\partial B_\parallel^* f_s}{\partial t} &+ \nabla \cdot \left(B_\parallel^* \left\{\mathbf{R}, H\right\} f_s\right) + \frac{\partial}{\partial v_{\parallel}}\left(B_\parallel^* \left\{v_{\parallel}, H\right\} f_s\right) \nonumber\\
    & = B_\parallel^* C\left(f_s\right) + B_\parallel^* S_s.
    \label{eq:gk_eq}
\end{align}
In Eq.~\eqref{eq:gk_eq}, $f_s$ is the gyrocenter distribution function for species $s$, $H = m_s v_\parallel^2/2 + \mu B + q_s \phi$ is the Hamiltonian of species $s$ with mass $m_s$, charge $q_s$, parallel velocity $v_\parallel$, magnetic moment $\mu$, magnetic field strength $B$, and electrostatic potential $\phi$.
We also define $B_\parallel^* = \mathbf{b} \cdot (\mathbf{B} + m_s v_\parallel / q_s \nabla \times \mathbf{b})$, the Jacobian of the coordinate transformation from particle to gyrocenter coordinates, and $\{\cdot, \cdot\}$ the Poisson bracket.
The collision operator, $C[f_s]$, is a Dougherty operator \citep{Dougherty1964,Francisquez2022improvedDougherty}, which conserves particles, momentum and energy, and relaxes the distribution function to a Maxwellian at a rate given by the collision frequency $\nu_{ss'}$ between species $s$ and $s'$, which is computed self-consistently from the local plasma parameters \citep{Francisquez2022improvedDougherty,hoffmann2026fullypredictivegyrokineticfullf}.
The limitations associated with this simplified model are discussed in previous work \citep{Dougherty1964,Francisquez2022improvedDougherty} and recalled in Appendix~\ref{sec:appendix_collisions}.
The source term, $S_s$, injects particles and heat, and is described in detail in \cite{hoffmann2026fullypredictivegyrokineticfullf} and Appendix~\ref{sec:appendix_setup}.

Finite-Larmor-radius corrections remain only in the polarization term of the quasi-neutrality equation, which is solved to obtain the electrostatic potential $\phi$,
\begin{equation}
    - \nabla \cdot \left(\epsilon_\text{pol} \nabla_{\perp}\phi\right) = \sum_s q_s \int d^3v f_s,
    \label{eq:quasineutrality}
\end{equation}
where $\epsilon_\text{pol} = \sum_s m_s n_{s0}/B_0^2$ is the polarization weight, with $n_{s0}$ the reference density of species $s$, $m_{s}$ its mass, $q_s$ its charge, and $B_0$ the reference magnetic field.
The use of the Boussinesq approximation, i.e.\ replacing the local weight $\sum_s m_s n_s/B^2$ by $\epsilon_\text{pol}$, on the left-hand side of Eq. \eqref{eq:quasineutrality} is a simplification suited to the L-mode conditions we focus on.
This approximation has been shown to modify the transport considerably in 2D models \citep{angus2014modelinglargeamplitude,Held2023BeyondApproximation} and when the gradients are large \citep{donnel2026fullfgkadaptive}, while it changes the turbulent transport only moderately in other cases \citep{ross2018effectboussinesq}.

The simulations use a field-line-following coordinate system \citep{Beer1995} in which the equilibrium magnetic flux surfaces are described by a Miller parameterization \citep{Miller1998NoncircularModel} with a quadratically varying Shafranov shift,
\begin{align}
    R(r, \theta) &= R_\text{axis} - \frac{\alpha_s r^2}{2 R_\text{axis}} + r \cos(\theta + \sin^{-1}\delta \sin\theta), \\
    Z(r, \theta) &= Z_\text{axis} + r \kappa \sin\theta,
\end{align}
where $R_\text{axis}$ is the major radius of the magnetic axis, $Z_\text{axis}$ its vertical position, $\delta$ the triangularity, $\kappa$ the elongation, and $\alpha_s$ the Shafranov shift parameter.
The safety factor profile is assumed to have a quadratic shape \citep{Riva2017PlasmaTurbulence}, $q(r) = q_\text{axis} + (q_\text{sep} - q_\text{axis}) (r/a)^2$, where $q_\text{axis}$ and $q_\text{sep}$ are the safety factor at the magnetic axis and at the separatrix, respectively, and $a$ is the minor radius of the separatrix.
This quadratic safety factor profile limits the normalized magnetic shear to values below 2, which is lower than typical values measured in the TCV boundary.
We nonetheless choose this simplified model for robustness across the large shaping scan.
Higher-order profiles, such as the cubic fits used in \cite{Hoffmann2025InvestigationSimulations}, must be carefully tuned to avoid nonphysical $q<1$ values and negative shear in the simulation domain.

\section{Simulation and scan setup}
\label{sec:scan_and_database}
The scan is designed to demonstrate that hundreds of unsupervised full-f boundary GK simulations can be run concurrently, as required by design optimization loops and by the training of data-driven models.
Large parameter scans can also provide physical insights when paired with a theoretical framework, as in previous local GK studies \cite{Merlo2023InterplayPlasmas,balestri2025interplaytriangularitymtm,Hoffmann2025InvestigationSimulations}, which we exemplify here with the effect of shaping on boundary turbulence.
The scan is composed of 256 simulations: 8 triangularity ($\delta$) values, 8 elongation ($\kappa$) values, and 4 power injection levels ($P_\text{in}$), i.e.,
$$
\delta \in \left\{\pm 0.15, \pm0.30, \pm0.45, \pm0.60\right\},
$$ 
$$
\kappa \in \left\{1.1, 1.2, 1.3, 1.4, 1.5, 1.6, 1.7, 1.8\right\},
$$ 
$$
P_\text{in} \in \left\{0.1, 0.5, 1.0, 5.0\right\} \text{ MW},
$$
respectively. 
The safety factor profile parameters are set to $q_\text{axis} = 1.2$ and $q_\text{sep} = 2.6$ for all simulations, which is consistent with the range of values observed in TCV \citep{Coda2022EnhancedTCV,duval2024tcvteam}.
The other equilibrium parameters are fixed to $R_\text{axis} = 0.87$ m, $Z_\text{axis} = 0.14$ m, $\alpha_s = 0.2$.
Reference magnetic field amplitude and plasma parameters are set to $B_0=1.129$ T,  
$n_0 = 2\times 10^{19}$ m$^{-3}$, and $T_0 = 100$ eV, yielding $\rho_{s0} = \sqrt{m_iT_0}/(eB_0) \simeq 1.3$ mm. The minor radius is $a = 0.25$ m at the outboard midplane.

The radial spatial domain spans from 0.04 m inside the separatrix to 0.08 m outside the separatrix at the outboard midplane, which corresponds to a range of normalized radius of $0.83 < r/a < 1.35$ and to a radial width of 100$\rho_{s0}$.
In the binormal direction, which can be associated with the toroidal direction at the outboard midplane, we set $L_y = 150 \rho_{s0}$, which is sufficient to capture the largest turbulent structures in the system \cite{hoffmann2026fullypredictivegyrokineticfullf}.
The parallel domain extends from $-\pi$ to $\pi$ in poloidal angle, terminating at the limiter.
In velocity space we use a non-uniform grid with $v_\parallel \in \left[-5 v_{ts}, 5 v_{ts}\right]$ and $\mu \in \left[0, 8 T_{s0}/B_0\right]$, where $v_{ts} = \sqrt{T_{s0}/m_s}$ is the thermal speed of species $s$ \citep{hoffmann2026fullypredictivegyrokineticfullf,Francisquez2026velocity_mappings}.

Particles are free to leave the domain through the radial boundaries and are re-injected with two adaptive sources \cite{hoffmann2026fullypredictivegyrokineticfullf} (see Appendix~\ref{sec:appendix_setup}), which maintains a constant number of particles in the system. The \textit{core source}, located at the inner radial boundary, injects a net power $P_\text{in}$ split evenly between the electrons and ions while replenishing the particle losses through the inner radial boundary. The \textit{recycling source}, located at the high-field side, models the neutral recycling process by adapting to the flux of particles leaving the domain through the wall and limiter, and re-injecting them as a $10$ eV Maxwellian, since recycled ions are much colder than the main ions \citep{wan2010naturalfuelingtokamak}.
The recycling source contributes less than $1\%$ of the total injected power.

The binormal direction boundaries are periodic, and the parallel boundaries use either twist-and-shift \citep{Beer1995,Ball2020,Francisquez2024ConservativeConditions} inside the separatrix ($r/a < 1.0$), or a conducting sheath model \citep{shi2019gkyltoklike} outside the separatrix ($r/a > 1.0$).
The potential is set to be zero at the wall and the limiter, which corresponds to a grounded vessel. At the inner radial boundary, we set $\phi=0$ for simplicity. This is not a physically accurate boundary condition, as the potential should be free to evolve up to the magnetic axis. However, a Neumann condition would also pose challenges, as it would require a radial force balance constraint on $E_r$ to be physically consistent. We leave the implementation of more accurate boundary conditions for future work and consider that the inner radial boundary is sufficiently far from the separatrix to not significantly affect the turbulence at the edge and SOL. The detailed form of these boundary conditions is given in Appendix~\ref{sec:appendix_setup}.
Consistent with the flux-driven nature of the code, the density and temperature profiles are free to evolve self-consistently in response to the sources, boundary losses, and turbulent transport throughout the domain, including at the innermost radial cells. The radial boundary condition on the distribution function is purely absorbing (see Appendix~\ref{sec:appendix_bc}), so no profile or gradient value is imposed there.

A Maxwellian distribution function is used as the initial condition for each simulation of the scan. The initial density profile is given by
\begin{equation}
    n(x) = n_0 \left\{1.0 + \tanh\left[-80\left(x - x_0\right) \right] \right\} + 0.001 n_0
\end{equation}
where $n_0 = 2\times10^{19}$ m$^{-3}$ and $x_0 = -0.03$ m. The initial electron and ion temperature profiles are given by
\begin{equation}
    T_e(x) = T_{e0} \left\{0.8 + 0.7\tanh\left[-80\left(x - x_0\right)\right]\right\},
\end{equation}
and
\begin{equation}
    T_i(x) = T_{i0} \left\{0.7 + 0.5\tanh\left[-30\left(x - x_0\right)\right]\right\},
\end{equation}
respectively. 

Each simulation uses a hybrid $p=1$ polynomial DG basis with $p=2$ in the parallel velocity direction \cite{Francisquez2026velocity_mappings}, yielding 48 degrees of freedom per cell.
The simulation domain is discretized in $(N_x, N_y, N_z) = (32, 24, 16)$ configuration space cells, and $N_{v_\parallel} = 12$ and $N_\mu = 8$ velocity space cells with a linear to quadratic mapping in $v_\parallel$ and a quadratic mapping in $\mu$. This resolution corresponds to $5.7\times10^7$ degrees of freedom in phase space.
The resolution considered here is not adjusted case by case, and is higher than the minimal resolution considered in \cite{hoffmann2026fullypredictivegyrokineticfullf} to provide a robust setup for the large number of simulations of the scan.
In units of the reference Larmor radius $\rho_{s0}\simeq1.3$~mm, the effective perpendicular spacing is $\Delta x \simeq 1.5\,\rho_{s0}$ (radial) and $\Delta y \simeq 3.1\,\rho_{s0}$ (binormal), so the ion-scale modes of interest ($k_\perp\rho_s\simeq0.4$, i.e. $\lambda_\perp\simeq16\,\rho_s$) are resolved with about ten points per wavelength radially and five binormally. Similar to local GK codes, fewer points are needed along the field-aligned coordinate $z$, where the turbulence is strongly elongated ($k_\parallel\ll k_\perp$).

The 256 simulations are run on the Perlmutter supercomputer at NERSC in one 128-node batch, with two NVIDIA A100 GPUs per \gkyl{} instance.
The input parameters for the scan, simulations and details about the version of \gkyl{} used are provided at \url{https://github.com/gkeyllorg/gkyl-paper-inp/tree/master/2026_PRL_tcv_miller_scan}.
Each simulation autonomously restarts until it reaches at least 1 ms of simulated time.
Since a change in geometry and input power affects the magnetic geometry and the turbulent dynamics, the average simulation time step varies across the scan, with a minimum of 2.2 ns and a maximum of 4.5 ns, the largest time steps being obtained at low elongation and weakly negative triangularity.
The total computational cost is $17{,}000$ GPU node hours for a wall-clock time of 5.4 days.

In the following section, we verify that each simulation reaches a quasi-stationary state, give a qualitative overview of the turbulent boundary, and then analyze the effect of shaping and power injection on the profiles and gradients across the scan.

\section{Confinement properties}
\label{sec:results}

\subsection{Saturation of the turbulence and quasi-steady state}
\label{sec:steady_state}

\begin{figure}[t]
    \centering
    \includegraphics[width=1.0\linewidth]{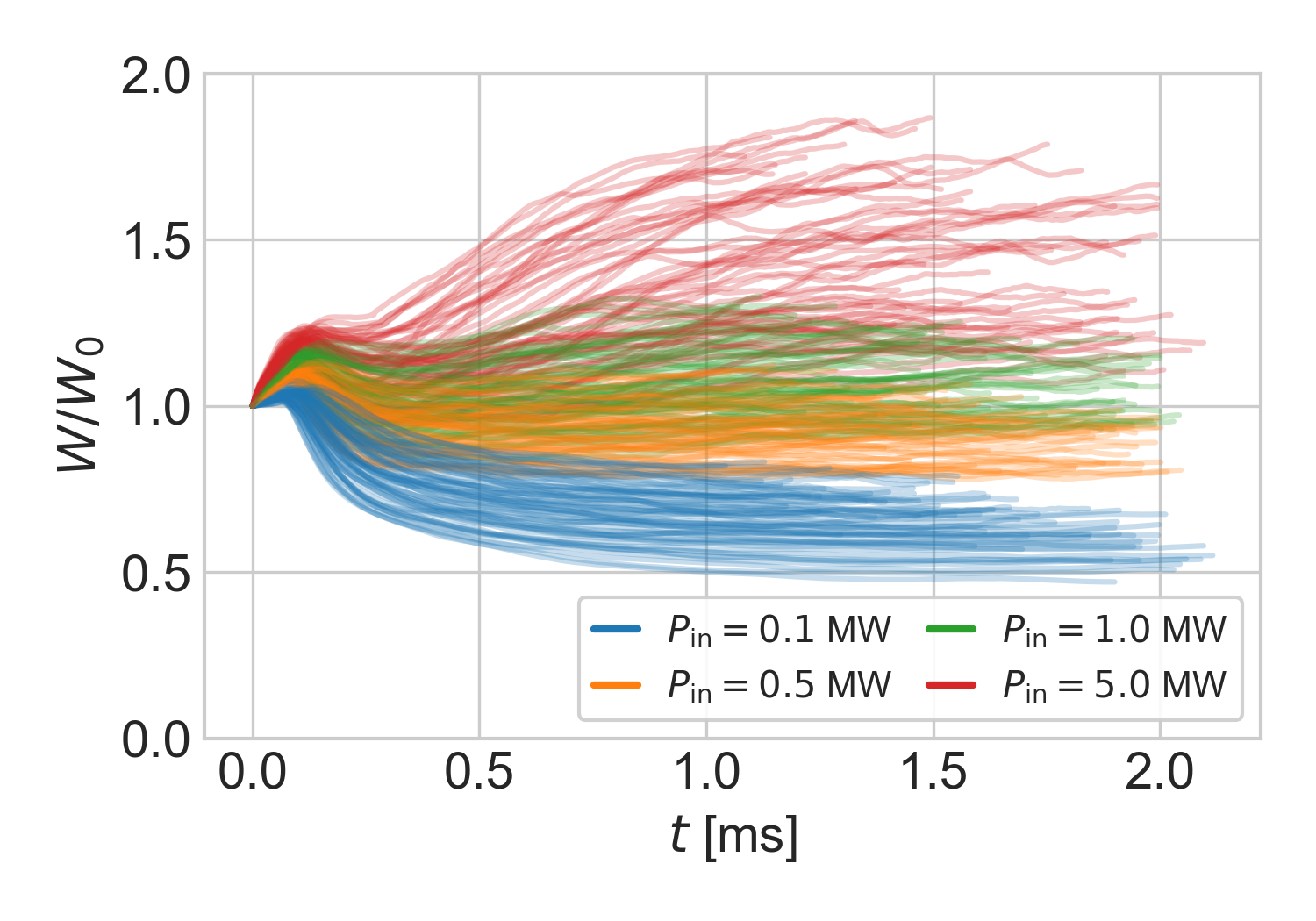}
    \caption{Time evolution of the particle energy, $W=\sum_s \int H_s f_s \mathrm{d}^3v \mathrm{d}^3x$ with $H_s$ the Hamiltonian of species $s$, normalized by the initial energy $W_0$.}
    \label{fig:total_energy_timetrace}
\end{figure}
\begin{figure*}[t]
    \centering
    \includegraphics[width=1.0\textwidth]{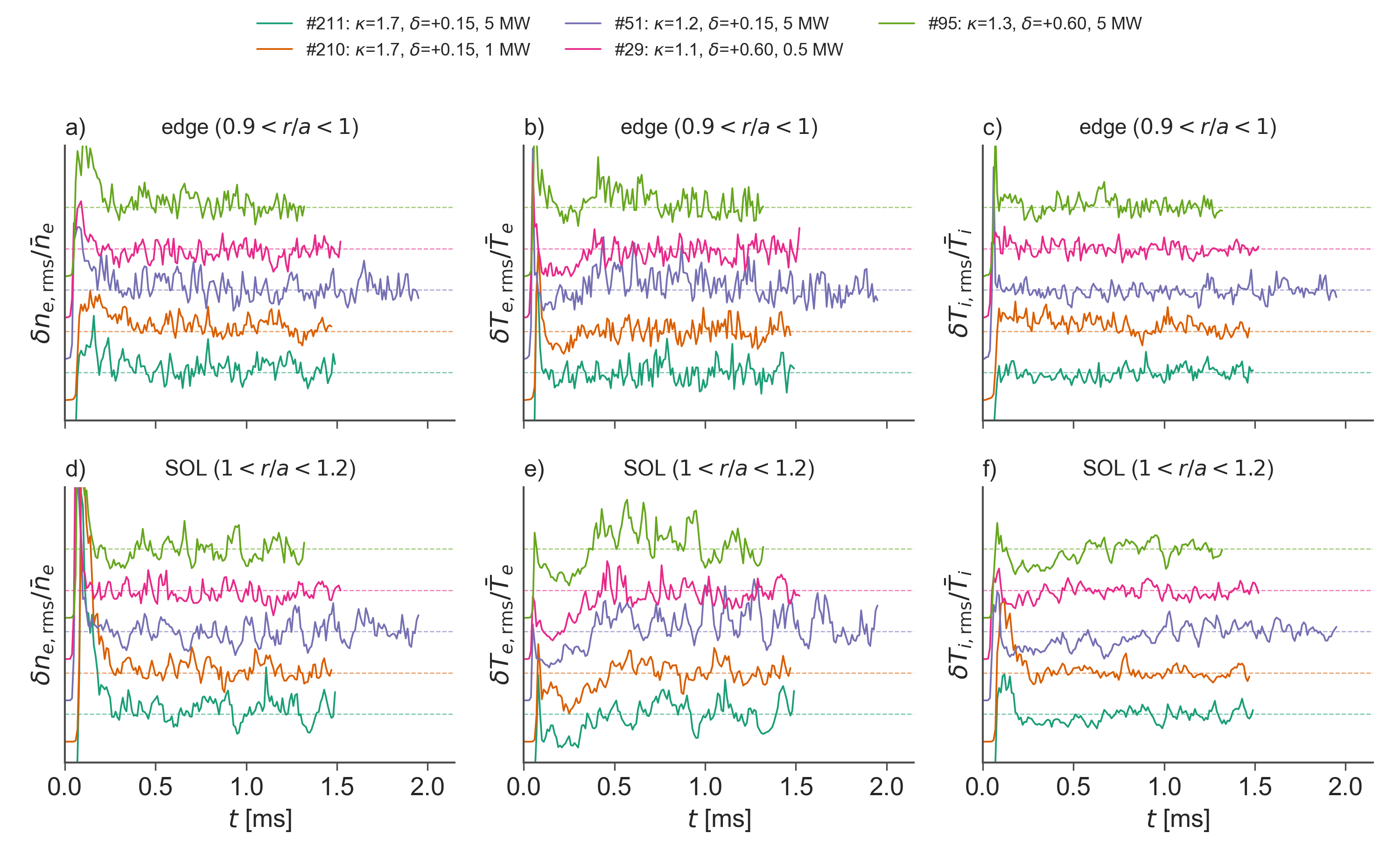}
    \caption{Relative fluctuation level $\delta X_\text{rms}/\bar X$, Eq.~\eqref{eq:fluct_level}, of the electron density (a, d), electron temperature (b, e) and ion temperature (c, f) in the edge ($0.9<r/a<1$, a--c) and the SOL ($1<r/a<1.2$, d--f), for the five simulations whose total energy varies the most over the last $200~\mu$s of the run. Each curve is normalized by its mean over the last $200~\mu$s (dashed line) and shifted vertically for readability.}
    \label{fig:dnrms_timetrace}
\end{figure*}

We verify that a quasi-steady state is reached in each simulation by monitoring the time evolution of the total energy of the system.
Figure~\ref{fig:total_energy_timetrace} shows that our simulation set encompasses both systems that relax to a lower energy state and systems that reach a higher energy state.
Despite the diversity in the final energy state, a shared transient structure is observed across all simulations.
In the first phase ($t \lesssim 0.1$ ms), the total energy increases monotonically as the system evolves from its initial condition.
Around $t \sim 0.2$ ms, the onset of turbulent transport counters the power injection and causes the energy to peak and begin to saturate or decrease depending on the input power level.
A second transition occurs near $t \sim 0.25$ ms, where the development of global $\ExB$ flows further modifies the turbulent transport.
At higher power injection the energy resumes its increase, while at lower power injection the energy loss rate is reduced.
Although these phases do not follow the same trend in energy across the scan, they occur at similar times in all simulations, reflecting a shared feedback timescale between the edge and SOL regions.
Because the relevant benchmark for this saturation is the slowest transport timescale of the simulated region rather than the turbulence correlation time, we estimate the energy confinement time of the simulation domain as $\tau_E = W / P_\text{loss}$, where $W$ is the total particle energy in the domain and $P_\text{loss}$ the power flowing out through the outer radial boundary and the two limiter ends.
We obtain an ensemble median $\tau_E \simeq 0.86$~ms, with individual values between $0.58$ and $1.6$~ms and a weak dependence on the injected power ($0.93$~ms at $0.1$~MW against $0.84$~ms at $5$~MW).
The simulations, which run between $1.04$ and $2.11$~ms, therefore cover $1.9$ energy confinement times of the simulated volume in the median, between $0.8$ and $3.3$ across the ensemble, with only ten simulations below one confinement time.
This is sufficient for the volume-averaged energy to saturate, as Fig.~\ref{fig:total_energy_timetrace} shows, but it remains a small number of confinement times, and a subset of the simulations, mostly at the highest injected power, still show residual total energy variations at the end of the run.
These residual variations are most likely tied to the slowest processes, in particular the mean-flow (neoclassical and zonal-flow) equilibration and the radial confinement of the closed region, which are of the order of a millisecond and are thus partially resolved.
We quantify this by analyzing the saturation of the local plasma energy.
We evaluate the time variation of the Hamiltonian energy integrated over the flux surface and over each radial cell, i.e.
\begin{equation}
    \dot W_i = \sum_s \iint \mathrm{d}z\, \mathrm{d}y \int_{x_i}^{x_{i+1}} \frac{\partial}{\partial t} \int H_s f_s\, \mathrm{d}^3v\, \mathrm{d}x,
\end{equation}
and average the absolute value over the radial cells, $\delta \dot W = N_x^{-1}\sum_{i=1}^{N_x} |\dot W_i|$, which yields an ensemble median of $7\%$ of the injected power, with $13\%$ at the 90th percentile.

We verify turbulence saturation with the relative fluctuation level of the density and temperatures,
\begin{equation}
    \frac{\delta X_\text{rms}}{\bar X} = \left\langle \left(\frac{\delta X}{\bar X}\right)^2 \right\rangle_r^{1/2},
    \label{eq:fluct_level}
\end{equation}
where $\delta X = X - \bar X$ is the local fluctuation, $\bar X = \frac{1}{L_y}\int_0^{L_y} X\, \mathrm{d}y$ is the binormal average, and $\langle\cdot\rangle_r$ the volume average over the region $r$, either the edge ($0.9<r/a<1$) or the SOL ($1<r/a<1.2$).
Figure~\ref{fig:dnrms_timetrace} shows the relative fluctuation levels of the five simulations whose total energy varies the most over the last $200~\mu$s, which we identify as the least converged cases of Fig.~\ref{fig:total_energy_timetrace}.
The density and temperature fluctuation levels are stabilized within the last $0.4$ ms of each simulation.
Taking a $200~\mu$s earlier time average changes the fluctuation levels of individual simulations by $4\%$ in the median and $12\%$ at the 90th percentile, without systematic drift, which is small compared with the variation of the fluctuation levels across the scan.
While this analysis indicates that subdominant local equilibration processes are still ongoing in some of the simulations, this work relies on trends across 256 simulations rather than on individual points.
The ensemble is therefore sufficiently converged for the analysis of the impact of shaping and power injection on the turbulent transport in the edge and SOL regions.

In the following, all quantities are averaged over the binormal direction and over the last $200~\mu$s of each simulation, the window over which the fluctuation levels of Fig.~\ref{fig:dnrms_timetrace} are stationary, and evaluated at the outboard midplane ($z=0$) at $r/a = 0.9$ (edge), $1.0$ (separatrix) and $1.2$ (SOL).
The sensitivity to the time average is assessed by repeating the analysis over the last $100~\mu$s only, which changes the ensemble medians of the profiles and gradients by less than $1\%$ and those of the transport coefficients by less than $5\%$, and over the independent window spanning the preceding $200~\mu$s, which changes the ensemble medians by less than $3\%$.
Neither affects any of the trends discussed below.

\subsection{Qualitative overview of the turbulent boundary}
\label{sec:overview}

\begin{figure*}[ht]
    \centering
    \includegraphics[width=0.7\textwidth]{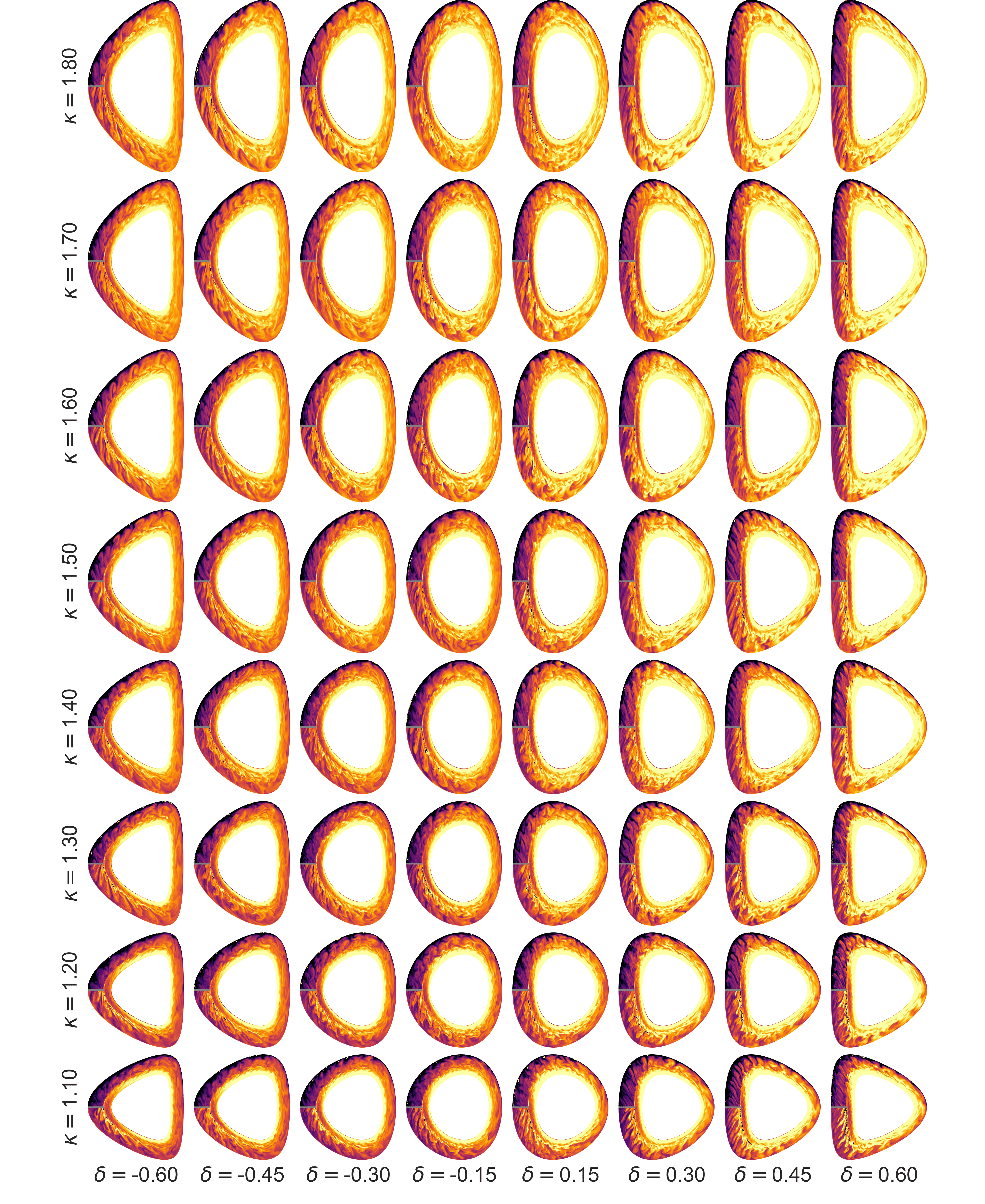}
    \caption{Poloidal cut of $T_i$ at $t=1$ ms for all shapes considered with $P_\text{in}=0.5$ MW. Each panel uses the same linear color mapping, between 0 and 220 eV.}
    \label{fig:poloidal_plot_array}
\end{figure*}
We present in Fig.~\ref{fig:poloidal_plot_array} a snapshot of the ion temperature on a poloidal cross section for all geometries at $P_\text{in} = 0.5$ MW and $t=1$ ms.
The color map is kept constant across configurations, allowing a visual comparison of the impact of shaping on the turbulent temperature structures.
Each configuration presents a $\nabla B$ drift pointing downwards, which is responsible for a higher temperature at the limiter bottom than at the top.
The hot plasma in the closed-field-line region is first transported radially outwards by $\ExB$ flows on the low-field side, where turbulence is more intense due to bad curvature.
Once the plasma reaches the SOL, the transport is mostly parallel and the plasma cools down through parallel heat convection to the limiter.
Radial transport is also present in the SOL and is an interplay between residual $\ExB$ flows, blob dynamics and trapped-particle orbits \citep{shi2019gkyltoklike,bernard2023neutralblobdynamics}.
In the vicinity of the limiter, the ion parallel flow accelerates towards the sheath and develops a strong radial gradient across the separatrix.
This configuration is favorable to parallel-velocity-gradient instabilities and its detailed analysis is left for future work \citep{garbet1999parallelkhinstability,fenzi1999densityfluctuations}.
Higher temperatures are observed at the low-field side of the positive triangularity configurations in Fig.~\ref{fig:poloidal_plot_array}, for both closed and open field line regions.
A more quantitative analysis is required to disentangle the effect of shaping on the turbulence and on the parallel transport in the SOL.

Localized artifacts are visible in the temperature field near the top of the domain and in the wake of the limiter corner, together with pixel-scale gradients near the outermost radius.
These are properties of the temperature diagnostic and not of the solution.
The temperature is obtained by a weak discontinuous Galerkin division of the pressure moment by the density, which is ill-conditioned in the few far-SOL cells where the density collapses intermittently below $10^{-2}$ of its separatrix value, while the density and energy-density moments remain smooth in the same cells.
The affected cells carry less than $2\times10^{-4}$ of the particle and energy content of the domain, their number does not grow in time, and the pixel-scale gradients are the sub-cell slopes of the $p=1$ DG representation rendered by the plotting interpolation.

\subsection{Shaping and power injection effects on profiles}
\label{sec:sep_prof_grad}

\begin{table}[ht]
    \centering
    \caption{Average and standard deviation of the separatrix conditions and SOL heat-flux decay length over the simulation ensemble.}
    \label{tab:sep_conditions}
    \begin{ruledtabular}
    \begin{tabular}{lcccc}
    $P_\text{in}$ [MW] & $n_e^\text{sep}$ [$10^{18}$ m$^{-3}$] & $T_e^\text{sep}$ [eV] & $T_i^\text{sep}$ [eV] & $\lambda_q$ [mm] \\
    \hline
    0.1 & $9.9 \pm 0.7$ & $89 \pm 25$  & $107 \pm 7$  & $3.9 \pm 0.9$ \\
    0.5 & $9.6 \pm 0.6$ & $114 \pm 19$ & $160 \pm 12$ & $4.2 \pm 1.1$ \\
    1.0 & $9.4 \pm 0.7$ & $130 \pm 18$ & $183 \pm 18$ & $4.0 \pm 1.0$ \\
    5.0 & $9.2 \pm 0.9$ & $167 \pm 23$ & $229 \pm 34$ & $3.6 \pm 1.0$ \\
    \end{tabular}
    \end{ruledtabular}
\end{table}

We first report the ensemble averages and variations of key plasma quantities at the separatrix in Table~\ref{tab:sep_conditions}.
The full ensemble statistics of these and of the other quantities discussed in this section, together with their marginal dependence on each scan parameter, are presented in Appendix~\ref{sec:appendix_statistics}.
The separatrix density stays close to $10^{19}$ m$^{-3}$ across the scan, while both ion and electron separatrix temperatures roughly double from the lowest to the highest injected power.
We also report the SOL heat-flux decay length $\lambda_q$, obtained by fitting an exponential to the binormally averaged parallel heat flux at the limiter over the first quarter of the SOL.
The ensemble median is $\lambda_q\simeq4.0$~mm, with individual values between $1.9$ and $8.5$~mm, a range compatible with L-mode measurements in TCV \citep{Nespoli2017lambdaqtcv}.
While a reduction of $\lambda_q$ at negative triangularity is reported experimentally in diverted L-mode plasmas \citep{faitsch2018lambdaqtriangularityTCV}, we find no significant dependence in this limited-geometry setup.
This can be due to different shaping effects between limited and diverted configurations, the experimental scaling being measured with an X-point, or to a missing physics mechanism in the simulations, such as a more realistic neutral model.

\begin{figure}[ht]
    \centering
    \includegraphics[width=1.0\linewidth]{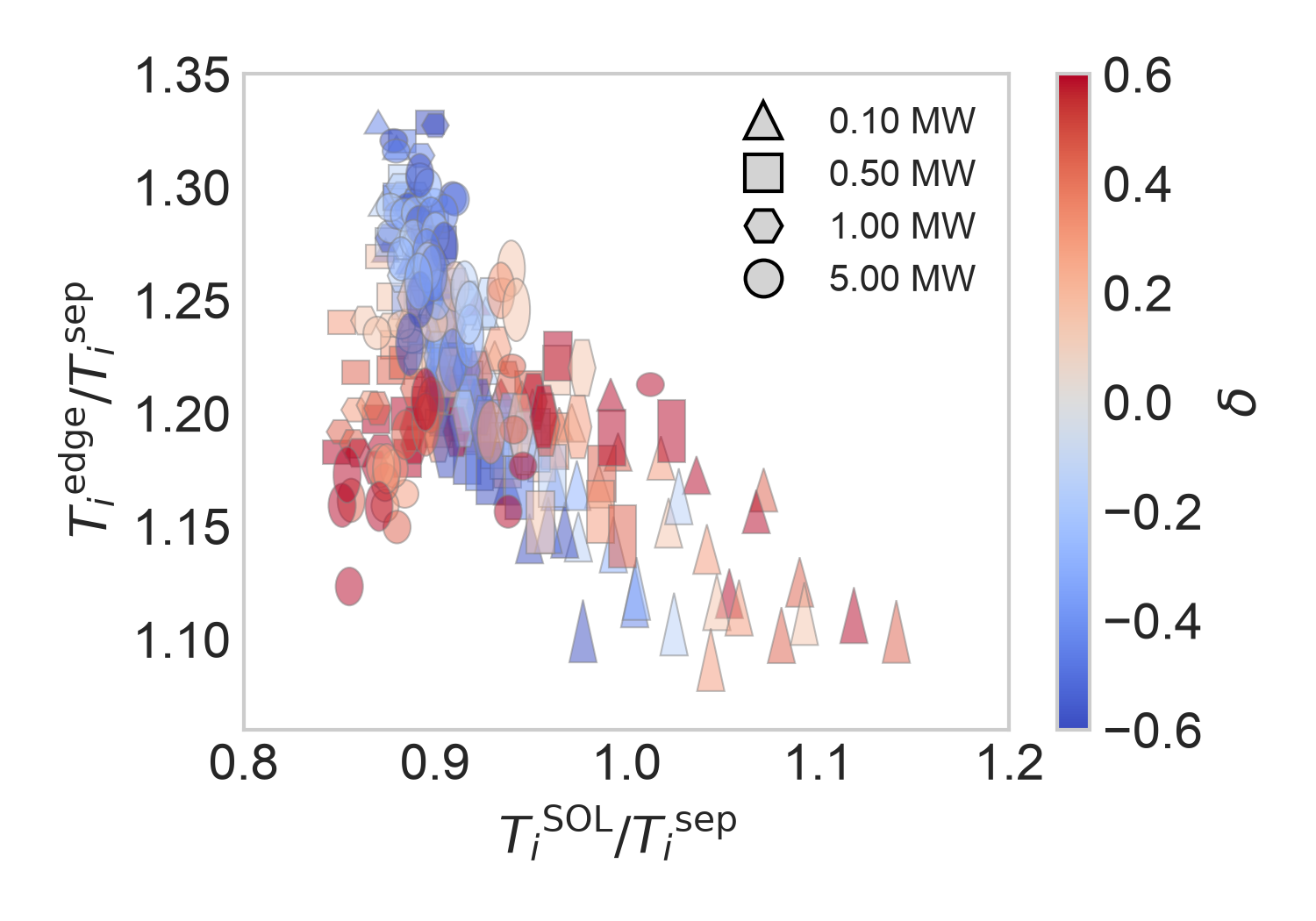}
    \caption{Edge ion temperature $T_i^\text{edge}$ vs SOL ion temperature $T_i^\text{SOL}$, both normalized by the separatrix temperature $T_i^\text{sep}$, for all simulations. The marker height encodes the elongation $\kappa$.}
    \label{fig:Tiedge_vs_Tisol}
\end{figure}
Power injection improves the ion energy confinement of the boundary, and more so at negative triangularity.
We analyze the ion temperatures at the edge and in the SOL, $T_{i,\text{edge}}$ and $T_{i,\text{SOL}}$, both normalized by the ion temperature at the separatrix, $T_{i,\text{sep}}$, to remove the global temperature increase driven by the varying input power.
Good confinement corresponds to a high $T_{i,\text{edge}}/T_{i,\text{sep}}$ and a low $T_{i,\text{SOL}}/T_{i,\text{sep}}$, and reflects a steep temperature gradient in the closed-field-line region.
Figure~\ref{fig:Tiedge_vs_Tisol} shows that higher power injection globally raises $T_{i,\text{edge}}/T_{i,\text{sep}}$ and lowers $T_{i,\text{SOL}}/T_{i,\text{sep}}$, indicating improved confinement, analogous to the L-H transition \citep{Gohil1994,Wagner2007,zholobenko2026grillixLHtransition}.
The improvement is more pronounced for negative triangularity, where increasing power simultaneously raises the edge temperature and lowers the SOL temperature.
In positive triangularity, the improvement is primarily a reduction of the relative SOL temperature, which can even exceed the separatrix temperature at the lowest power level (Sec.~\ref{sec:trapped_ions}).

\begin{figure}[ht]
    \centering
    \includegraphics[width=1.0\linewidth]{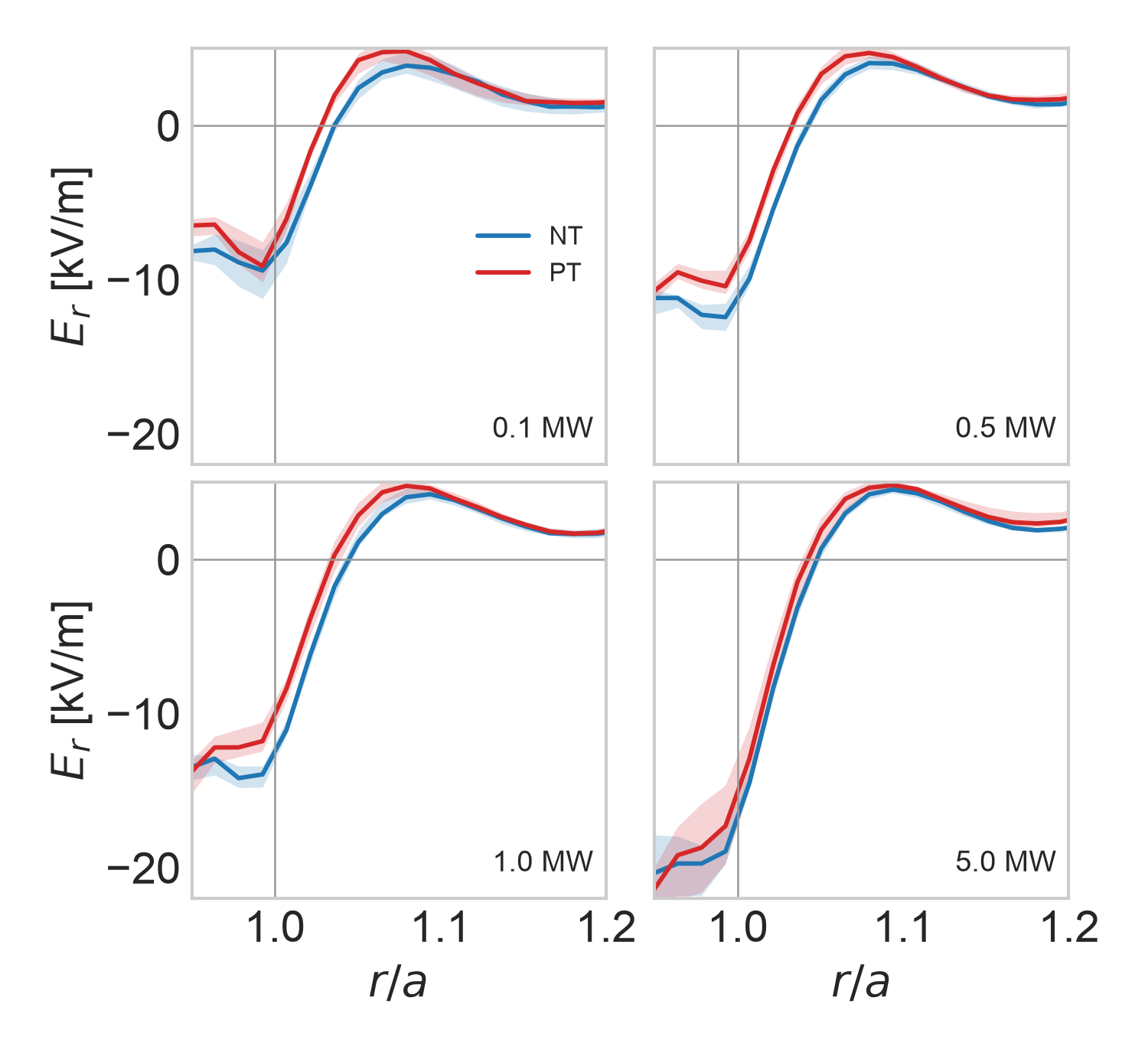}
    \caption{Radial profiles of the radial electric field $E_r$ at the outboard midplane, averaged over the binormal direction and over the last $200~\mu$s, as ensemble medians (lines) and interquartile ranges (bands) over the 32 negative (NT) and 32 positive (PT) triangularity simulations at each injected power.}
    \label{fig:Er_profiles}
\end{figure}

\begin{figure}[ht]
    \centering
    \includegraphics[width=1.0\linewidth]{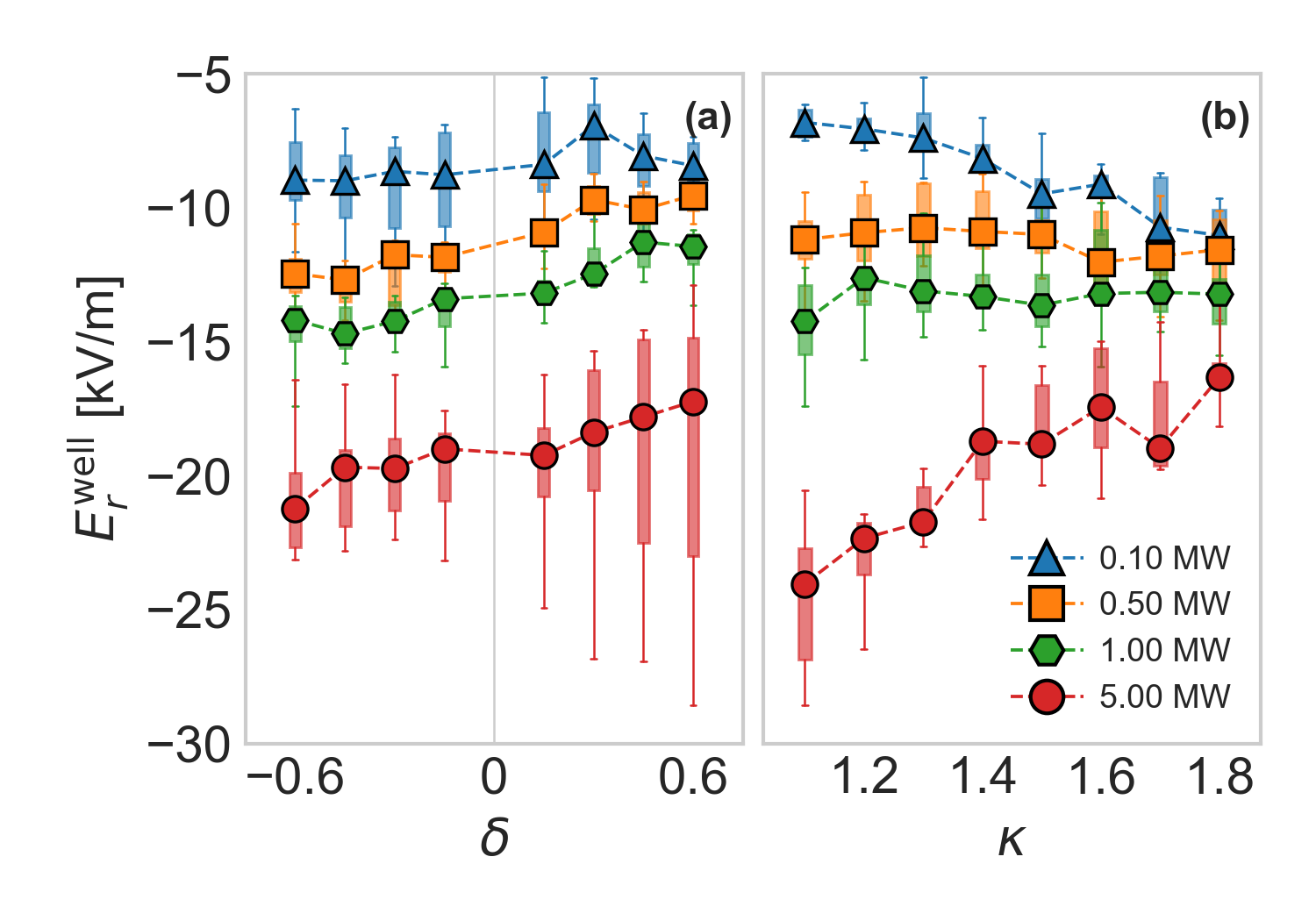}
    \caption{Radial electric field well, i.e. $E_r$ at the outboard midplane at $r/a=0.98$, as a function of the triangularity $\delta$ (a) and of the elongation $\kappa$ (b). Each box spans the eight elongations at a given $\delta$ (a), and the eight triangularities at a given $\kappa$ (b) and constant $P_\text{in}$. The markers indicate the medians of each power level.}
    \label{fig:Er_well}
\end{figure}

As in the L-H transition, the confinement improvement comes with a deepening of the radial electric field well inside the separatrix.
Figure~\ref{fig:Er_profiles} shows the radial profiles of $E_r = -\partial_x \phi$ averaged over time and binormal direction at the outboard midplane, as ensemble medians over the negative and positive triangularity simulations at each power.
The well is deeper in negative triangularity at intermediate power, as measured in TCV \citep{Rienacker2026ErNTPTtcv}, but more complex trends are observed at low and high power.
The analysis of the electric field well $E^\text{well}_r$, which we define as $E_r$ evaluated at $r/a = 0.98$, reveals a strong dependence on both triangularity and elongation, as shown in Fig.~\ref{fig:Er_well}.
The radial electric field well deepens when the power is increased, from $-8$ kV/m at $0.1$ MW to $-20$ kV/m at $5$ MW in the median, and with negative triangularity at intermediate power, $-12.3$ against $-10.1$ kV/m at $0.5$ MW and $-14.2$ against $-12.2$ kV/m at $1$ MW in the median, with a Spearman rank correlation of $0.8$ with $\delta$ at these two powers. The correlation with $\delta$ is much weaker at high and low power ($0.2$), where we find that elongation plays a larger role.
At $0.1$ MW, elongation deepens the well, from $-6.8$ to $-11.0$ kV/m between $\kappa=1.1$ and $1.8$ in the median, with a rank correlation of $-0.85$ between $E_r^\text{well}$ and $\kappa$, while at $5$ MW the well is shallower in elongated configurations, from $-24.0$ to $-16.3$ kV/m, with a rank correlation of $0.82$.
At $5$ MW, the well remains deeper in negative triangularity for $\kappa \geq 1.3$, with rank correlations between $0.4$ and $0.8$ with $\delta$, but the trend reverses at low elongation, with rank correlations of $-0.76$ and $-0.55$ at $\kappa = 1.1$ and $1.2$.
This reversal could be tested experimentally.

The $\ExB$ shearing rate $|\partial_x (E_r/B)|$ averaged over $0.94 < r/a < 0.99$ is nonetheless larger in positive triangularity at every power, from $2.4\times10^5$ against $1.8\times10^5$ s$^{-1}$ at $0.1$ MW to $6.4\times10^5$ against $2.8\times10^5$ s$^{-1}$ at $5$ MW in the median, because $E_r$ rises more steeply towards the separatrix in positive triangularity.
The edge temperature ratio is higher in negative triangularity at every power, e.g.\ $T_{i,\text{edge}}/T_{i,\text{sep}} = 1.28$ against $1.19$ at $5$ MW in the median, so the confinement improvement of negative triangularity is not associated with a stronger sheared flow.
These results indicate that this improvement is due, in our simulations, to a direct reduction of the instability drive, which is examined in Sec.~\ref{sec:instabilities}.

The simulated well can be qualitatively compared with the Doppler backscattering measurements of the edge $\ExB$ velocity in TCV L-mode plasmas \citep{Rienacker2025radialelectricfieldTCVLmode,Rienacker2026ErNTPTtcv}.
With $B \simeq 1.13$ T at the outboard midplane, the simulation well corresponds to a perpendicular velocity $v_\perp = E_r/B$ of $-7$ km/s at $0.1$ MW in the median, of the same order as the $-4$ km/s well of limited Ohmic discharges \citep{Rienacker2025radialelectricfieldTCVLmode}, and the well is deeper in negative than in positive triangularity in matched Ohmic and heated discharges, $-4$ against $-2$ km/s for $\delta \simeq \pm 0.3$ \citep{Rienacker2026ErNTPTtcv}, as in our simulations at intermediate power.
In TCV, electron cyclotron heating, which mainly heats the electrons, leaves the $v_\perp$ profile nearly unchanged.
With co-current neutral beam injection, the torque raises the core $v_\perp$ while the well does not follow, so that the well deepens relative to the core flow and the velocity shear increases with the injected power \citep{Rienacker2025radialelectricfieldTCVLmode}.
Our simulations show a similar deepening, with $v_\perp$ reaching $-17$ km/s at $5$ MW, and the near-separatrix shearing rate reported above also increases with power, although no momentum is injected.
A source injecting toroidal momentum is the natural test of this agreement, and a quantitative comparison is probably also limited by the inaccurate neoclassical effects of the Dougherty operator.

\begin{figure}[ht]
    \centering
    \includegraphics[width=1.0\linewidth]{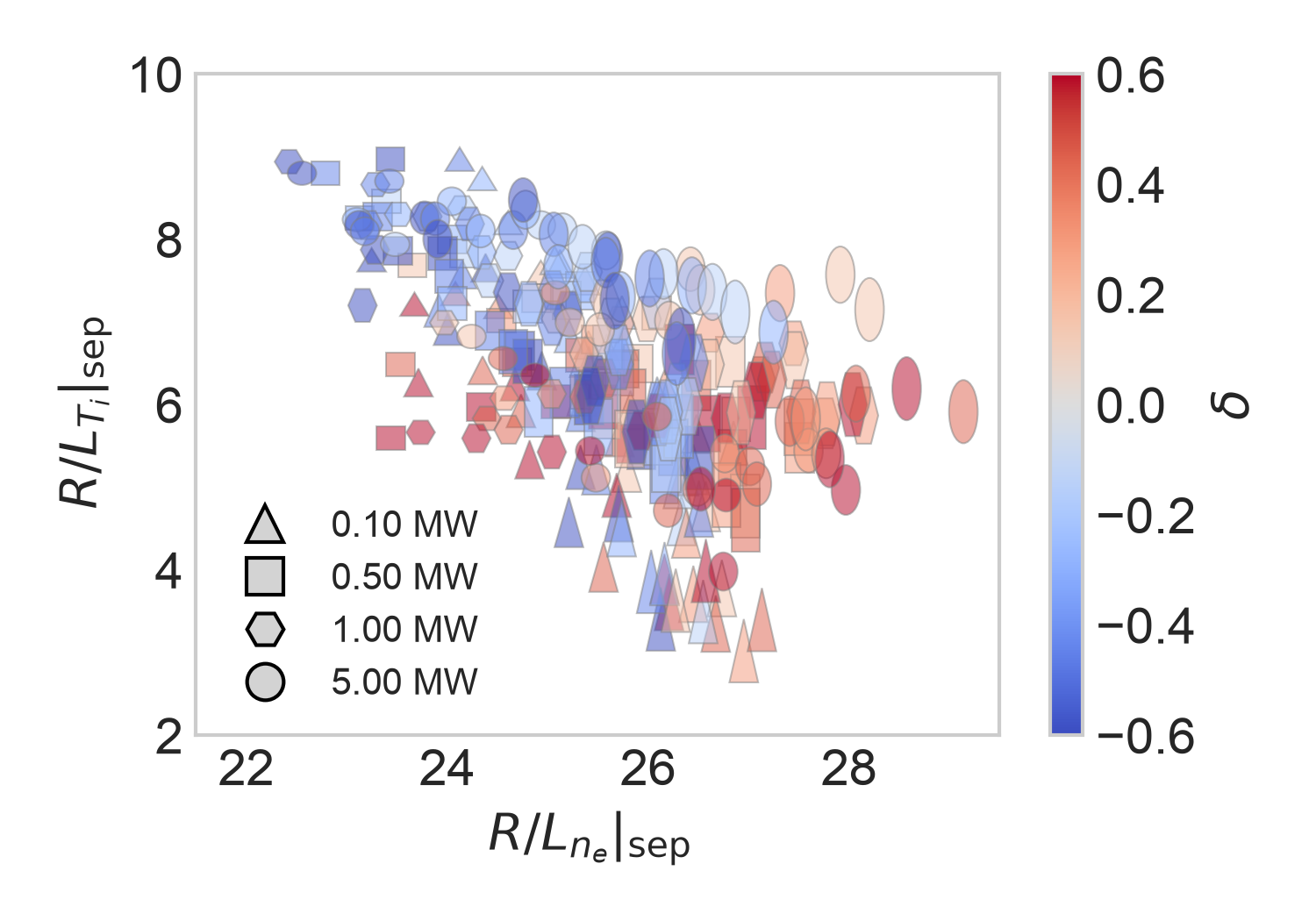}
    \caption{Normalized ion temperature gradient $R/L_{T_i}$ as a function of the normalized electron density gradient $R/L_{n_e}$, both evaluated between the edge ($r/a=0.9$) and the separatrix ($r/a=1.0$) following Eq. \eqref{eq:norm_gradient}, for all simulations. The marker height encodes the elongation $\kappa$.}
    \label{fig:gradient_analysis}
\end{figure}
We now turn to the normalized gradients of density and ion temperature just inside the separatrix, where a pedestal would form, which are key parameters for the stability of microinstabilities.
We estimate the gradients from the edge and separatrix values, $\partial_x X = (X_\text{sep} - X_\text{edge})/(0.1\,a)$, where $X$ is a plasma quantity of interest, and normalize them as
\begin{equation}
    R/L_X = -R_\text{axis}\,\frac{\partial_x X}{X_{\text{edge}}}.
    \label{eq:norm_gradient}
\end{equation}
Figure~\ref{fig:gradient_analysis} reports the normalized ion temperature gradient $R/L_{T_i}$ as a function of the normalized density gradient $R/L_{n_e}$ for all simulations, with $R/L_{T_i}$ between $3.0$ and $9.0$ across the scan.
Negative triangularity steepens the edge ion temperature profile more efficiently with power.
The median $R/L_{T_i}$ rises from $6.3$ at $0.1$ MW to $8.0$ at $5$ MW in negative triangularity, against $5.4$ to $5.8$ in positive triangularity.
A stronger sensitivity of the edge gradients to shaping is observed at higher triangularity.
The standard deviation of the normalized gradients increases with $|\delta|$, by about $15\%$ in $R/L_{T_i}$ and $40\%$ in $R/L_{n_e}$ between $|\delta| = 0.15$ and $0.60$.
\begin{figure}[ht]
    \centering
    \includegraphics[width=1.0\linewidth]{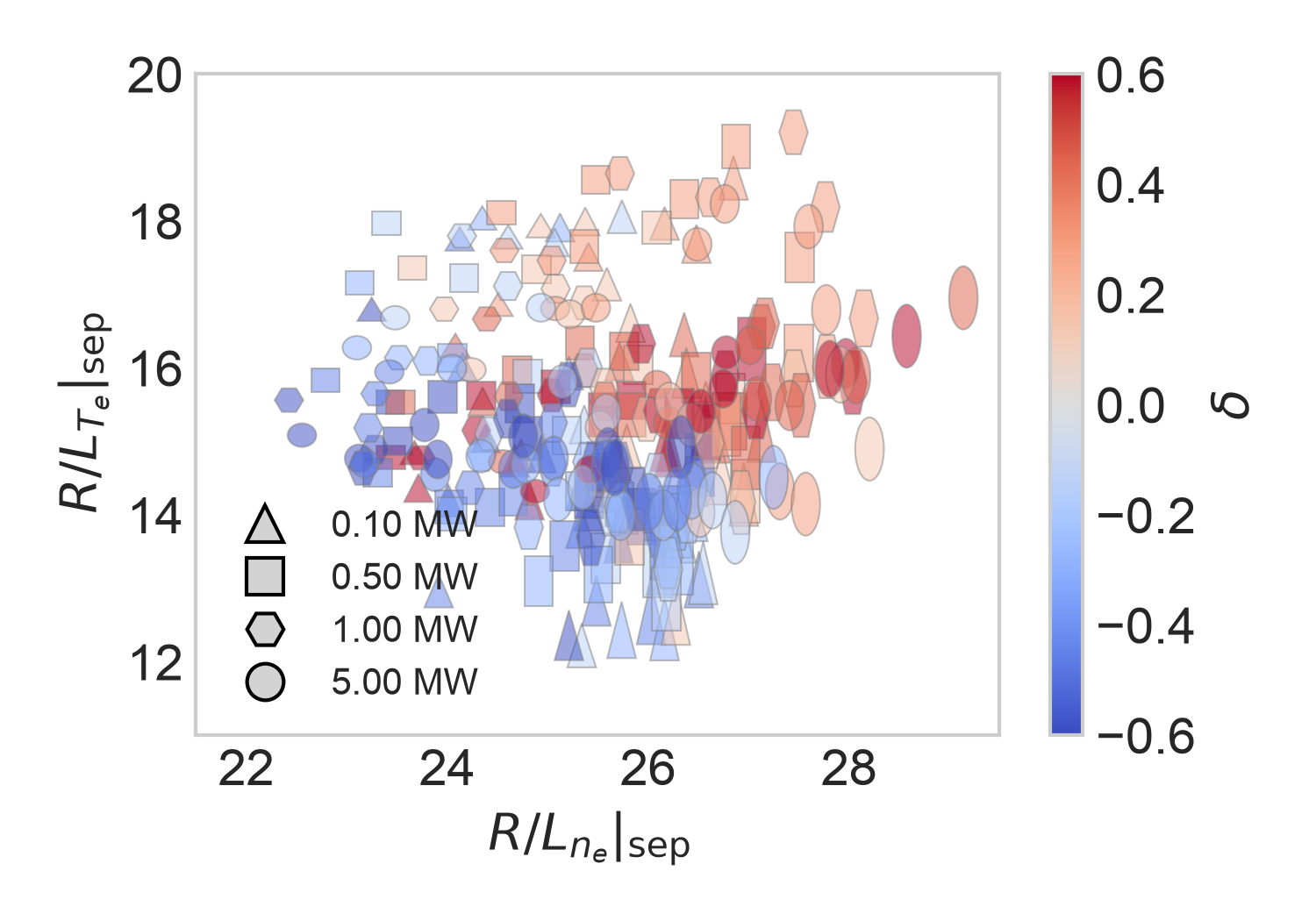}
    \caption{Normalized electron temperature gradient $R/L_{T_e}$ as a function of the normalized electron density gradient $R/L_{n_e}$, both evaluated between the edge and the separatrix following Eq.~\eqref{eq:norm_gradient}, for all simulations. The marker height encodes the elongation $\kappa$.}
    \label{fig:gradient_Te}
\end{figure}

The electron temperature gradient, see Fig.~\ref{fig:gradient_Te}, varies less than the ion one and is slightly larger in positive triangularity ($15.2$ at $0.1$~MW, $15.8$ at $5$~MW in the median) than in negative triangularity ($14.0$ and $14.7$).
The different response of the two temperature gradients to shaping may be attributed to the effect of triangularity on the microinstabilities \citep{Belli2008EffectsTurbulence,Merlo2019TurbulentTriangularity,Bernard2024PlasmaPlasmas,Hoffmann2025InvestigationSimulations,Balestri2024PhysicalPlasmas}, which we examine next.

\section{Microinstability identification and transport analysis}
\label{sec:instabilities}
It is established that a density gradient drives TEMs, while it can stabilize ITG modes \citep{Garbet2004PhysicsTokamaks,casati2008temperatureratioITGTEM}.
This stabilizing role is not universal, since at sufficiently large $R/L_n$ the density gradient can itself destabilize ITG modes, so the mapping from gradients to dominant instability is only indicative and motivates the dedicated linear and transport-coefficient analysis.

\subsection{Linear gyrokinetic simulations}
We perform electrostatic linear simulations with the local flux-tube GK code \gyacomo{} \citep{Hoffmann2023GyrokineticShift,Hoffmann2025InvestigationSimulations}.
The parameters of the linear simulations are extracted from the nonlinear \gkyl{} simulation data between the edge and the separatrix, i.e. the local magnetic geometry, the density and temperature gradients of Eq.~\eqref{eq:norm_gradient}, the electron-ion temperature ratio, and the collisionality.
The equilibrium $\ExB$ flow shear developed in the \gkyl{} simulations is not included in the linear simulations, which therefore give unsheared growth rates.

\begin{figure}[ht]
    \centering
    \includegraphics[width=1.0\linewidth]{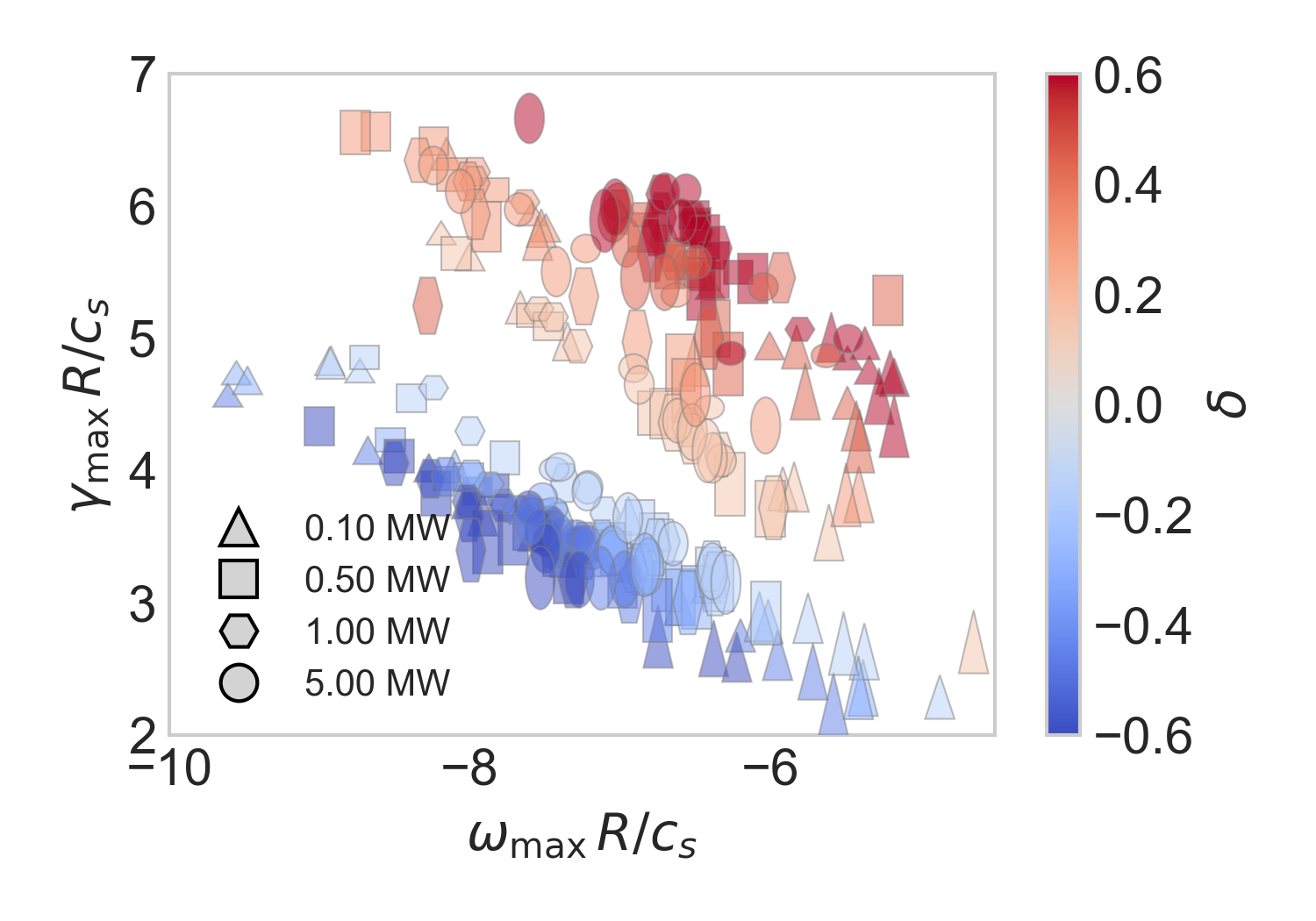}
    \caption{Linear growth rate $\gamma_\text{max}$ of the most unstable mode, and its frequency $\omega_\text{max}$ , for the local parameters of Fig.~\ref{fig:gradient_analysis}. The marker height encodes the elongation $\kappa$.}
    \label{fig:linear_gk}
\end{figure}

For each of the 256 sets of local parameters we scan the binormal wavenumber over $k_y\rho_s\in[0.2,1.0]$ and retain the most unstable mode at each $k_y$.
We verify the convergence of the growth rate $\gamma$ and frequency $\omega$ by doubling the parallel resolution ($N_z = 48$) or the velocity-space resolution (9 Hermite and 5 Laguerre moments instead of 5 and 3), which changes the maximal $\gamma$ by less than $3\%$.
The $k_y\rho_s \leq 0.1$ modes are found to be stable and thus excluded from the presented spectra.
Figure~\ref{fig:linear_gk} shows the growth rate of the most unstable mode, $\gamma_\text{max}$, as a function of its frequency, $\omega_\text{max}$, for all simulations of the scan.
We compute $\gamma_\text{max}$ with a quadratic interpolation around the spectrum maximum and $\omega_\text{max}$ by the same interpolation of $\omega$ at the wavenumber of the maximum.
The peak lies at $k_y\rho_s = 0.35$ in the median, between $0.30$ and $0.38$ for $80\%$ of the simulations.
All linear simulations predict that TEM instabilities dominate, as indicated by the negative frequency of the most unstable mode (electron diamagnetic direction in our convention).
The linear results of Fig.~\ref{fig:linear_gk} aggregate along nearly constant $\gamma/\omega$ lines that correlate with triangularity.
At $k_y\rho_s=0.4$, near the spectral peak, $|\gamma/\omega|$ increases monotonically from $0.47$ at $\delta=-0.60$ to $0.93$ at $\delta=+0.60$, with a spread below $6\%$ within each triangularity subset, the latter value indicating a strong instability drive far from marginal stability.
The increase of $|\gamma/\omega|$ with triangularity can be linked to the resonance condition of the mode with the magnetic drift frequency, which depends strongly on the triangularity \citep{Balestri2024PhysicalPlasmas}.
A general reduction of the growth rate with increasing elongation is observed due to flux expansion, from $4.6$ at $\kappa=1.1$ to $3.3$ at $\kappa=1.8$ in the median at $k_y\rho_s=0.4$, consistent with former works \citep{Belli2008EffectsTurbulence,Merlo2023InterplayPlasmas,Garbet2024shapingTEMeffect,Hoffmann2025InvestigationSimulations}.

\begin{figure}[ht]
    \centering
    \includegraphics[width=1.0\linewidth]{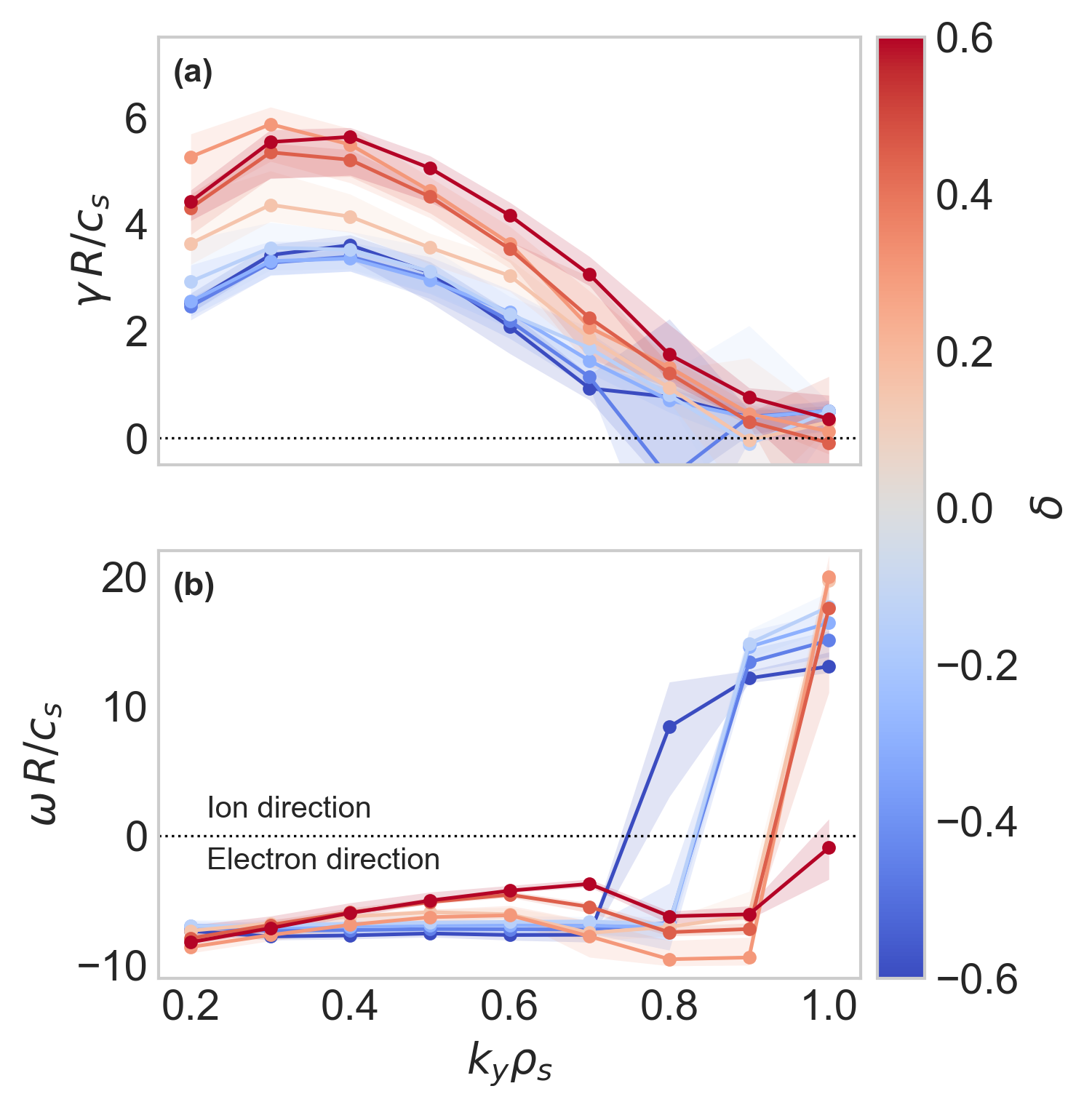}
    \caption{Growth rate (a) and frequency (b) of the most unstable mode in the linear \gyacomo{} simulations as a function of $k_y\rho_s$. Each curve is the median over the eight elongations and four power levels at a given triangularity, with the interquartile range shaded.}
    \label{fig:lin_spectra}
\end{figure}

Figure~\ref{fig:lin_spectra} shows the ensemble median of the growth rate and frequency spectra.
All unstable modes with $k_y\rho_s < 0.8$ have negative frequency, and present a larger growth rate in positive triangularity, $5.1$ against $3.4$ at $k_y\rho_s=0.4$ in the median, with a rank correlation of $0.77$ with $\delta$.
Among the local parameters, the TEM growth rate correlates most with the electron temperature gradient (rank correlation $0.79$) and with the temperature ratio $T_i/T_e$ ($0.68$), and only weakly with the density gradient ($0.18$).
The linear spectra also report a subdominant short-wavelength ion branch.
The frequency changes sign at $k_y\rho_s \geq 0.8$ in 227 of the 256 cases, revealing an ITG-like branch that coexists with the TEM but carries a growth rate an order of magnitude smaller, $\gamma R/c_s \simeq 0.4$ at $k_y\rho_s = 1.0$ in the ensemble median.
This branch responds to triangularity in the opposite way to the TEM.
In negative triangularity, the frequency reverses at $k_y\rho_s = 0.8$--$0.9$ and the ion branch is present in every simulation.
In positive triangularity, the reversal is pushed to $k_y\rho_s = 1.0$, and the growth rate of the ion branch decreases with $\delta$.
Opposite responses of the ITG and TEM branches to triangularity have been reported in local GK simulations, with a sign that depends on the magnetic shear \citep{Merlo2023InterplayPlasmas,Merlo2023OnTokamaks,Duff2022EffectTurbulence}, and the moderate magnetic shear of our quadratic safety factor profile (Sec.~\ref{sec:model}) may contribute to the stabilization of the ion branch in positive triangularity observed here. 

\subsection{Transport and instability fingerprint analysis}
\label{sec:transport}

We now investigate the nonlinear transport coefficients to corroborate the linear mode identification with nonlinear results.
Our analysis employs the transport channel parameter space to obtain information on the dominant instability, inspired by the fingerprint methodology \citep{Kotschenreuther_2019}.
We define the particle and heat diffusivities,
\begin{equation}
    D_s = -\frac{\Gamma_s}{\partial_x n_s}, \quad \chi_s = -\frac{Q_s}{n_s\partial_x T_s},
    \label{eq:flux_gradient_relations}
\end{equation}
respectively, where $\Gamma_s = \langle n_s v_{r}\rangle$ is the radial particle flux with the radial cross-field velocity $v_r$ computed from the $\ExB$ and $\nabla B$ drifts, and $Q_s$ the conductive radial heat flux, i.e. the part of the radial energy flux that is not carried by the particle flux.
We evaluate $Q_s$ directly as the correlation between the temperature and the turbulent particle flux,
\begin{equation}
    Q_s = \frac{3}{2}\left\langle \delta T_s\, \delta\!\left(n_s v_{r}\right)\right\rangle,
    \label{eq:conductive_heat_flux}
\end{equation}
where $\delta X = X - \overline{X}$ and the average is taken over the binormal direction and over the last $200~\mu$s at the outboard midplane.
The gradients $\partial_x X$ are the edge-to-separatrix differences defined above Eq.~\eqref{eq:norm_gradient}.

\begin{figure}[ht]
    \centering
    \includegraphics[width=1.0\linewidth]{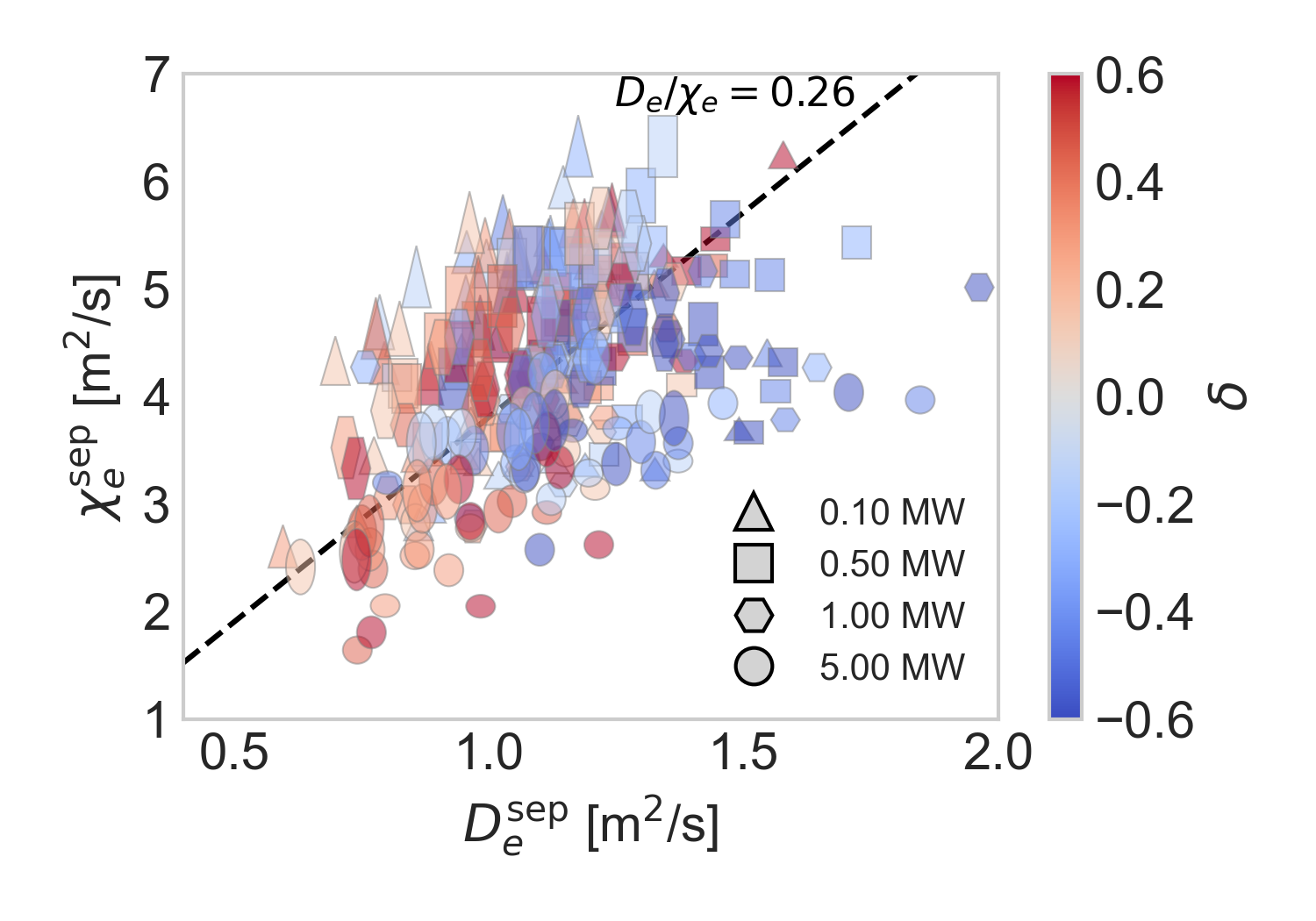}
    \caption{Separatrix electron heat diffusivity $\chi_e$ as a function of the electron particle diffusivity $D_e$ for all simulations. The marker height encodes the elongation $\kappa$.}
    \label{fig:De_vs_chie}
\end{figure}

Figure~\ref{fig:De_vs_chie} shows the separatrix electron heat diffusivity, $\chi_e$, as a function of the electron particle diffusivity, $D_e$, for the whole ensemble.
Both are positive in every simulation, with medians $D_e = 1.11$ m$^2$/s and $\chi_e = 4.24$ m$^2$/s.
They are positively correlated across the scan, with a ratio $D_e/\chi_e$ between $0.20$ and $0.36$ for $80\%$ of the simulations around a median of $0.26$, i.e.\ $\chi_e\simeq4\,D_e$.
From the exact variance decomposition of the balanced $8\times8\times4$ factorial design, $43\%$ of the $\chi_e$ variance is due to the injected power, against less than $10\%$ each for the shaping parameters.
In the median over all shapes, the electron heat diffusivity falls monotonically from $4.7$ m$^2$/s at $P_\text{in}=0.1$ MW to $3.3$ m$^2$/s at $5$ MW.
The electron energy transport is dominated by conduction, with a conductive to convective ratio, $Q_e/(\frac{3}{2}\overline{T}_e\Gamma_e)$, decreasing from $89\%$ at $0.1$ MW to $59\%$ at $5$ MW, a reduction of the transport with increasing power that is again reminiscent of the L-H transition.

\begin{figure}[ht]
    \centering
    \includegraphics[width=1.0\linewidth]{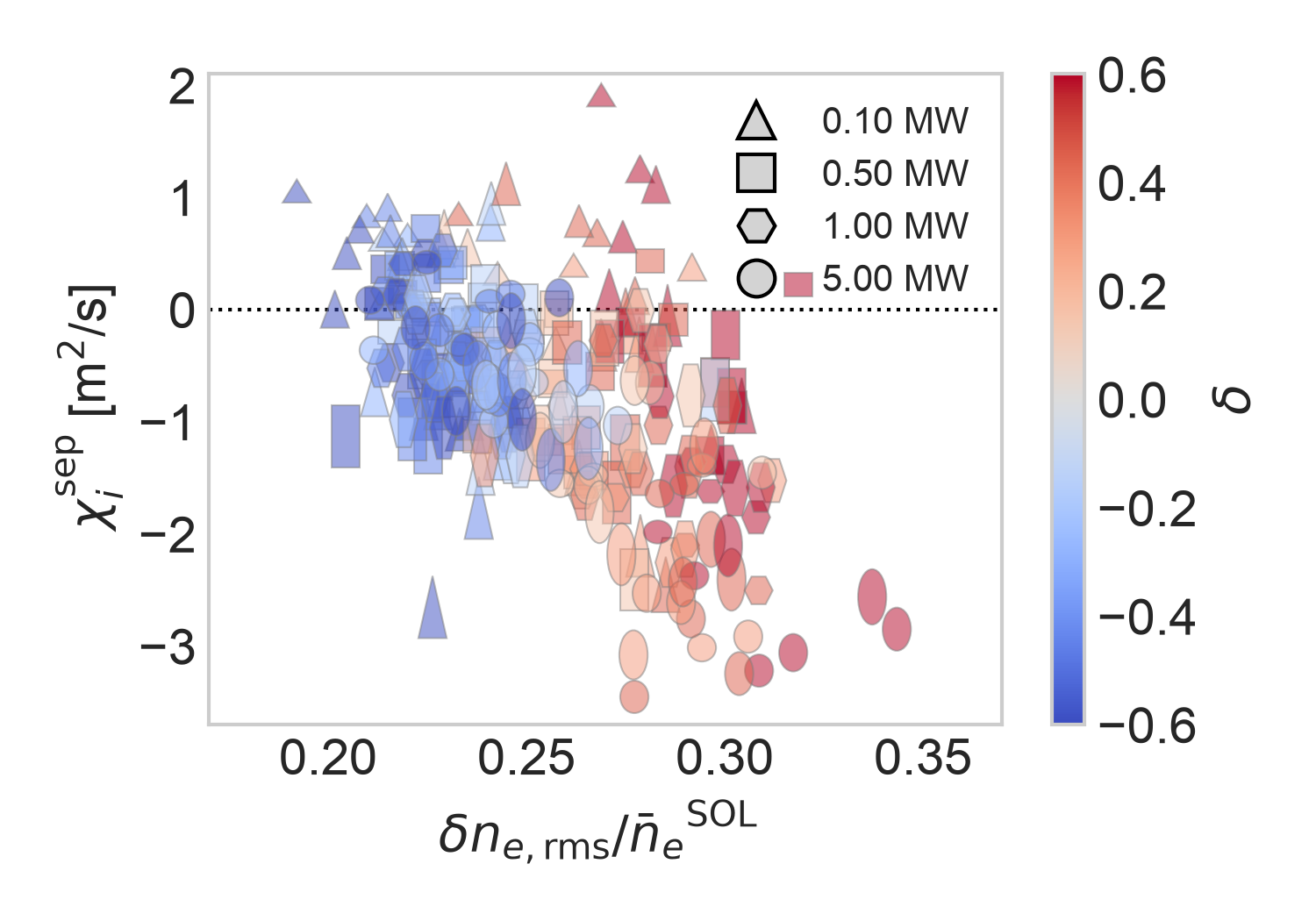}
    \caption{Separatrix ion heat diffusivity $\chi_i$ as a function of the relative density fluctuation level in the SOL ($1<r/a<1.2$), Eq.~\eqref{eq:fluct_level}, for all simulations. The marker height encodes the elongation $\kappa$.}
    \label{fig:chii_vs_dnrms}
\end{figure}

The ion transport channel behaves very differently from the electron one.
The ion particle diffusivity decreases monotonically with triangularity, from a subset median of $1.20$ m$^2$/s at $\delta=-0.60$ to $0.44$ m$^2$/s at $\delta=+0.60$ (Table~\ref{tab:ens_delta}).
The ion heat diffusivity, shown in Fig.~\ref{fig:chii_vs_dnrms}, is an order of magnitude below the electron one, with a median $\chi_i = -0.58$ m$^2$/s, and it is negative in $77\%$ of the simulations, most strongly at $5$ MW and positive triangularity, i.e. the cases with the lowest normalized SOL ion temperature in Fig.~\ref{fig:Tiedge_vs_Tisol}.
The conductive ion heat flux is also small compared with the convective one, $|Q_i| < 0.07 \times \frac{3}{2}\overline{T}_i\Gamma_i$ in the median at every power, so the ion energy transport is essentially convective.
Such an inward ion heat flux is supported by quasilinear theory, in which the magnetic drift resonance produces an inward contribution to the ion energy flux for both ITG and TEM turbulence \citep{wang2011heatpinchtheory}.
Figure~\ref{fig:chii_vs_dnrms} plots $\chi_i$ against the relative density fluctuation level in the SOL, $\delta n_{e,\text{rms}}/\bar n_e$ of Eq.~\eqref{eq:fluct_level}.
Negative triangularity reduces the SOL fluctuation level, from a median of $0.28$ in positive triangularity to $0.23$ in negative triangularity, monotonically in $\delta$ (Table~\ref{tab:ens_delta}), which is consistent with the reduced SOL density fluctuations reported in TCV negative-triangularity L-mode plasmas \citep{han2021suppressionfluctuationsNT}.
The highest fluctuation levels coincide with the lowest $\chi_i$ values, with a Spearman rank correlation of $-0.55$ over the ensemble that strengthens with power, up to $-0.83$ at $5$ MW.

The conductive heat flux is therefore carried almost entirely by the electrons, the fingerprint of an electron-driven mode \citep{Kotschenreuther_2019}.
Since neither electron temperature gradient modes nor micro-tearing modes are representable in the present electrostatic, long-wavelength model, the nonlinear fluxes point toward the TEM identified with \gyacomo{}.

\section{Trapped ions and recycling}
\label{sec:trapped_ions}

\begin{figure*}[ht]
    \centering
    \includegraphics[width=0.7\linewidth]{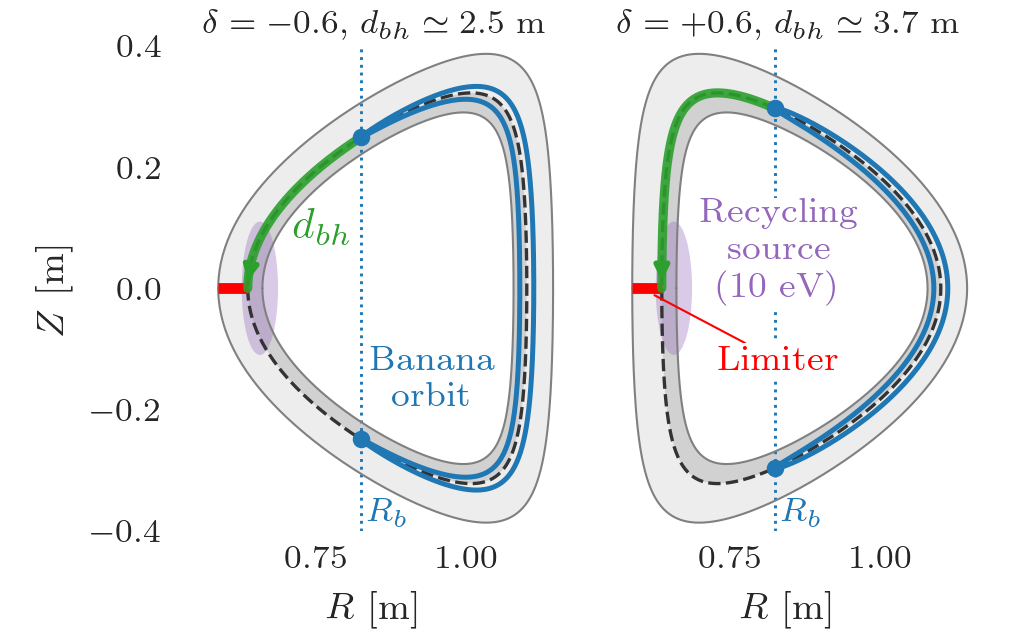}
    \caption{Illustration of the field-line distance $d_{bh}$ from the banana turning point $R_b$ to the recycling source on the high-field side, for $\delta=-0.6$ (left) and $\delta=+0.6$ (right) at $\kappa=1.4$. The green line is the poloidal projection of the path, the shaded rings are the simulated edge and SOL, and the banana width is exaggerated for legibility.}
    \label{fig:trapped_ion_sketch}
\end{figure*}

At low power, positive triangularity flattens the ion temperature profile across the separatrix.
At $P_\text{in}=0.1$ MW, the median $T_{i,\text{SOL}}/T_{i,\text{sep}}$ is $1.01$ in positive triangularity against $0.93$ in negative triangularity (Fig.~\ref{fig:Tiedge_vs_Tisol}), and $T_{i,\text{SOL}}$ exceeds $T_{i,\text{sep}}$ in 20 of the 64 simulations at this power, against 2 of the 192 simulations at higher power.
Flat SOL ion temperature profiles are measured in tokamaks \citep{kocan2011ionenergiesboundary,allan2016iontemperatureMAST}, and recycling is known to be a key factor in setting the SOL ion temperature \citep{wan2011pinchcoldions,Coroado2022ABoundary}.
We propose that this flattening results from the interplay between the trapped ions and the cold recycling source, and gather here the elements of the ensemble that support this interpretation.

The ion distribution is anisotropic at the separatrix, with a median $T_{i,\perp}/T_{i,\parallel} = 1.45$ over the ensemble and $1.49$ at $0.1$ MW, where $T_{i,\perp}$ and $T_{i,\parallel}$ denote the ion perpendicular and parallel temperatures.
This anisotropy indicates a large population of trapped ions, which remain on banana orbits crossing the separatrix while the passing ions are lost along the open field lines.
Ion orbits crossing the separatrix are known to shape the boundary, in particular through ion orbit loss and its effect on the radial electric field \citep{shaing1989bifurcationLH,chang2002xtransport,degrassie2015ionorbitloss}.
The ion collisionality is consistent with this picture.
From the separatrix conditions of Table~\ref{tab:sep_conditions}, the ion-ion mean free path is $17$ m at $0.1$ MW and $80$ m at $5$ MW, and the normalized collisionality $\nu_i^* = \nu_{ii}\, q_\text{sep} R_\text{axis}/(\epsilon^{3/2} v_{ti})$, with $\epsilon = a/R_\text{axis}$, decreases from $0.9$ to $0.2$, so the ions are at the boundary of the banana regime at low power and inside it at high power.
The anisotropy also explains the sign of the ion conductive flux.
Since $T_s=(T_{s,\parallel}+2T_{s,\perp})/3$ holds pointwise, the conductive heat flux of Eq.~\eqref{eq:conductive_heat_flux} splits exactly into a parallel and a perpendicular contribution,
\begin{equation}
    Q_s = \tfrac{1}{2}\left\langle \delta T_{s,\parallel}\,\delta\Gamma_s\right\rangle + \left\langle \delta T_{s,\perp}\,\delta\Gamma_s\right\rangle,
    \label{eq:qcond_split}
\end{equation}
with $\Gamma_s = n_s v_r$ the instantaneous particle flux.
For the ions, the two terms of Eq.~\eqref{eq:qcond_split} have opposite signs in $99\%$ of the ensemble, the parallel contribution being positive, i.e. down-gradient, and the perpendicular one negative.
At $r/a=0.9$, where the ions are instead sub-thermal in the perpendicular direction ($T_{i,\perp}/T_{i,\parallel}=0.82$), both contributions are positive in every simulation and the local ion heat diffusivity, evaluated with the same gradient, is $\chi_i=+1.7$~m$^2$/s in the ensemble median.
The negative ion heat diffusivity is therefore a feature localized at the separatrix, carried by the perpendicular temperature fluctuations of a trapped-rich population.

Since no direct heating is applied in the SOL, its ion temperature is set by the cross-field transport from the edge, the collisional heat exchange with the recycled ions, and the parallel losses to the limiter.
The only cold population in the model is the recycling source, a Maxwellian at $10$~eV localized on the high-field side (Appendix~\ref{sec:appendix_sources}), whose mean free path is of the order of $10$ cm.
Thus, the cooling due to recycling reaches the trapped ions by heat conduction along the field line, and becomes less efficient as the field-line distance $d_{bh}$ between the banana turning point and the high-field side increases.
For the median anisotropy at $0.1$ MW, which shows no trend with triangularity ($1.44$ to $1.56$ across the eight $\delta$ values), the turning point lies at the major radius $R_b = R_{\text{sep}}\, 2 T_{i,\perp} / (T_{i,\parallel} + 2T_{i,\perp}) \simeq 0.82$ m, where $R_\text{sep}$ is the major radius of the separatrix at the outboard midplane, which defines a poloidal turning angle $\theta_b$ by $R(a,\theta_b)=R_b$.
The field-line distance from this point to the high-field side at $\theta=\pi$ is
\begin{equation}
    d_{bh} = \int_{\theta_b}^{\pi} \sqrt{\left(\partial_\theta R\right)^2 + \left(\partial_\theta Z\right)^2 + R^2\left(\partial_\theta \varphi\right)^2}\; \mathrm{d}\theta,
    \label{eq:arc_length}
\end{equation}
where $\partial_\theta \varphi$ is the local field-line pitch.
Evaluating Eq.~\eqref{eq:arc_length} at the separatrix ($q=q_\text{sep}=2.6$) gives $d_{bh}\simeq2.5$~m for $\delta=-0.60$ against $3.7$~m for $\delta=+0.60$, as sketched in Fig.~\ref{fig:trapped_ion_sketch}, and $d_{bh}$ increases monotonically with $\delta$.
Since $\kappa$ does not vary radially in our Miller parameterization, $d_{bh}$ varies by less than $1\%$ across the elongation scan at fixed $q_\text{sep}$.
In this picture, positive triangularity lengthens the path between the trapped ions and the cold recycled population, weakens their collisional coupling, and leaves the SOL ions hotter than in negative triangularity.
The ensemble results are consistent with this interpretation.
At $P_\text{in}=0.1$~MW and fixed elongation, $T_{i,\text{SOL}}/T_{i,\text{sep}}$ correlates with $\delta$ with a partial Spearman rank correlation of $0.90$, and the correlation drops below $0.2$ in absolute value from $0.5$~MW on, consistent with a cooling effect that weakens as the injected power increases.
The connection length from the outboard midplane to the limiter, which sets the parallel losses, cannot explain this ordering, as it is shorter in positive triangularity ($6.4$ m at $\delta=+0.60$ against $7.2$ m at $\delta=-0.60$) and would thus favor a colder SOL there.

However, $d_{bh}$ is a function of $\delta$ alone, so this correlation cannot by itself separate the distance mechanism from other triangularity effects.
In particular, triangularity may also act on the SOL ion temperature through the turbulent inward pinch of the recycled ions \citep{wan2011pinchcoldions}, enhanced by the higher fluctuation level of positive triangularity (Fig.~\ref{fig:chii_vs_dnrms}).
The two levers are not separable in the ensemble, where the SOL fluctuation level and $\delta$ have a rank correlation of $0.82$ at $0.1$~MW.
The SOL ion temperature also rises with elongation, from $T_{i,\text{SOL}}/T_{i,\text{sep}} = 0.89$ at $\kappa=1.1$ to $1.06$ at $\kappa=1.8$ at $0.1$~MW on average over $\delta$, a trend that persists up to $1$~MW, where triangularity no longer correlates with $T_{i,\text{SOL}}/T_{i,\text{sep}}$, and that $d_{bh}$ cannot explain.
Finally, the energy exchange with the recycled population is a crude proxy for the neutral physics, where ionization and charge exchange provide a temperature-dependent cooling in the SOL that the present model lacks \citep{Coroado2022ABoundary,bernard2023neutralblobdynamics}.

\section{Conclusions}
\label{sec:conclusions}

This work demonstrates that high-throughput, unsupervised full-f boundary GK simulations are feasible, with 256 simulations of TCV-inspired equilibria completed in a single campaign of 5.4 days and $17{,}000$ GPU node hours.
The scan reveals a power-dependent effect of triangularity on confinement.
At high power, negative triangularity steepens the edge ion temperature gradient, consistent with the enhanced confinement reported for this shape in TCV \citep{Coda2022EnhancedTCV} and DIII-D \citep{Austin2019AchievementTokamak}.
Since the $\ExB$ shearing rate near the separatrix is larger in positive triangularity, we attribute this improvement to a weaker instability drive rather than to a stronger sheared flow.
At low power, positive triangularity flattens the ion temperature profile across the separatrix, for which we propose a trapped-ion mechanism based on the field-line distance between the banana turning points and the cold recycling source.
Linear GK simulations and the transport coefficients extracted from the turbulent fluxes consistently indicate a TEM-dominated regime, as in previous works \citep{Merlo2023InterplayPlasmas,Balestri2024PhysicalPlasmas,Hoffmann2025InvestigationSimulations}, in which the conductive heat flux at the separatrix is carried by the electrons and the ion energy transport is convective.
Without tuned transport coefficients or experimentally informed inputs, the ensemble reproduces the range of the SOL heat-flux decay length measured in TCV, the reduced SOL density fluctuations of negative triangularity, and the deeper radial electric field well of negative triangularity at intermediate power, which supports the predictive value of the simulation framework.

These trends remain to be confirmed with a more realistic model, including electromagnetic fluctuations \citep{Halpern2013IdealLayer,balestri2025interplaytriangularitymtm}, a collision operator that captures the neoclassical damping of the poloidal flow \citep{Hoffmann2023GyrokineticOperators,donnel2026fullfgkadaptive}, a self-consistent neutral model \citep{Bernard2022KineticModelingNeutral,Coroado2022ABoundary}, and a diverted geometry with radially varying shaping.
The neutral model is also the natural test of the trapped-ion mechanism, whose geometric argument applies to other cold regions on closed flux surfaces, such as the X-point radiator \citep{bernert2023xpointradiatingregime}.

The resulting 75 TB simulation ensemble, with configuration-space snapshots every 2 $\mu$s and full distribution functions every 20 $\mu$s, is publicly available \citep{hoffmann2026tcvmillerscandata}.
Together with the ensemble statistics of Appendix~\ref{sec:appendix_statistics}, it can serve as a reference for local gyrokinetic and integrated modelling \citep{Snyder2009PedestalPrediction,meneghini2024fuse}, as a benchmark for reduced transport models, and as training data for surrogate models of the boundary.
The throughput demonstrated here puts optimization and uncertainty-quantification workflows \citep{Highcock2018gkoptimisation,Kim2023OptimizationStellarators} within reach of full-f boundary GK.

\begin{acknowledgments}
The authors thank J. Juno, F. Parra Diaz, A. Balestri, S. Brunner and J. Ball for helpful discussions.
This work is supported by a DOE Distinguished Scientist award, the CEDA SciDAC project and other PPPL projects via DOE Contract Number DE-AC02-09CH11466 for the Princeton Plasma Physics Laboratory.
This research used resources of the National Energy Research Scientific Computing Center (NERSC), a Department of Energy User Facility.
An AI tool (Claude, Anthropic) was used for proofreading and verification. All conceptualization of simulations, data analysis, and original drafting were conducted by the authors.
\end{acknowledgments}

\section*{Data Availability}

The simulation ensemble analyzed in this work, comprising approximately 75 TB of configuration-space snapshots (every 2 $\mu$s) and full distribution functions (every 20 $\mu$s) for the 256 simulations, is openly available \citep{hoffmann2026tcvmillerscandata}.
The input files for the scan, together with the version of \gkyl{} used to produce it, are available at \url{https://github.com/gkeyllorg/gkyl-paper-inp/tree/master/2026_PRL_tcv_miller_scan}.

\appendix

\section{Additional details on the simulation setup}
\label{sec:appendix_setup}

To make the manuscript self-contained, this appendix gives the detailed form of the collision operator, the adaptive source terms, and the boundary conditions. The complete framework is developed and described extensively in \cite{hoffmann2026fullypredictivegyrokineticfullf,Francisquez2026velocity_mappings}.

\subsection{Collision operator}
\label{sec:appendix_collisions}

The collision operator $C[f_s]$ appearing in the GK equation, Eq.~\eqref{eq:gk_eq}, is a multi-species version of the long-wavelength GK Lenard--Bernstein--Dougherty operator \citep{Lenard1958,Dougherty1964,Francisquez2022improvedDougherty},
\begin{equation}
    \label{eq:appendix_collisions}
    J_s C[f_s] = \sum_r \nu_{sr} \grad_v \cdot \left[ \left( \bm v - \bm u_{sr} \right) J_s f_s + v_{t,sr}^2 \grad_v J_s f_s \right],
\end{equation}
where $J_s$ is the phase-space Jacobian, $\nu_{sr}$ the collision frequency of species $s$ with species $r$, and $\bm u_{sr}$ and $v_{t,sr}$ are cross-species primitive moments constructed to enforce exact momentum and energy conservation \citep{Francisquez2022improvedDougherty}. The collision frequencies are space- and time-dependent,
\begin{equation}
    \nu_{sr}(\bm R,t) = \nu_{sr0}\, \frac{n_r(\bm R,t)}{n_{r0}} \left[\frac{v_{ts0}^2 + v_{tr0}^2}{v_{ts}^2(\bm R,t) + v_{tr}^2(\bm R,t)}\right]^{3/2},
\end{equation}
with the reference collision frequency
\begin{equation}
    \nu_{sr0} = \frac{1}{m_s}\left(\frac{1}{m_s}+\frac{1}{m_r}\right)\frac{q_s^2q_r^2\log\Lambda_{sr}}{3(2\pi)^{3/2}\epsilon_0^2}\frac{n_{r0}}{\left(v_{ts0}^2+v_{tr0}^2\right)^{3/2}},
\end{equation}
and $\log\Lambda_{sr}\simeq14$ the Coulomb logarithm. This operator conserves particles, momentum, and energy, and relaxes the distribution function to a local Maxwellian at the collision rate $\nu_{sr}$. Because the modelled collision frequency does not reproduce the $\sim v^{-3}$ velocity dependence of the full Fokker-Planck operator, this model is not expected to capture neoclassical transport quantitatively. In particular, effects such as neoclassical impurity thermal screening are outside its scope. It nonetheless provides a conservative, self-consistent relaxation to a local Maxwellian, which is the dominant role of collisions for the boundary turbulence considered here.

\subsection{Adaptive source terms}
\label{sec:appendix_sources}
Gkeyll is a flux-driven code. The plasma profile is therefore not prescribed but emerges self-consistently from the interplay of the sources, boundary losses, and turbulent transport. The source terms are implemented in a fully conservative manner, so that the total number of particles and energy in the domain changes only through the boundary fluxes.
A distinguishing feature of this framework is an adaptive sourcing scheme that dynamically adjusts the particle and energy injection in response to the boundary losses measured by \gkyl{} at each Runge--Kutta sub-step \citep{hoffmann2026fullypredictivegyrokineticfullf}. The total volumetric source $S_s$ entering the GK equation, Eq.~\eqref{eq:gk_eq}, for species $s$ is written as a sum of sub-sources,
\begin{equation}
    S_s = \sum_{k=1}^{N_S} S^k_s,
\end{equation}
each a non-drifting Maxwellian in velocity space and a Gaussian in configuration space,
\begin{equation}
    S^k_s\left[\mathbf{R}, \vpar, \mu; \Gamma^k_s, T^k_s\right] = \Gamma^k_s\, \eta^k_s(\bm R)\, M^k_s\left[\vpar, \mu; T^k_s\right],
\end{equation}
where the semicolon separates the coordinate dependence from the source parameters, i.e. the particle injection rate $\Gamma^k_s$ and temperature $T^k_s$. The normalized spatial profile is a Gaussian in the radial and parallel coordinates,

\begin{equation}
    \eta^k_s(\mathbf{R}) = C^k_s \prod_{i=x,z} \exp\left[-\frac{(x_i - \xi_{i,s,k})^2}{2\sigma_{i,s,k}^2}\right],
\end{equation}
with $C^k_s$ enforcing $\int \eta^k_s\, d\mathbf{R}= 1$, and the velocity-space profile is the Maxwellian
\begin{equation}
    M^k_s\left[\vpar, \mu; T^k_s\right] = \left(\frac{m_s}{2\pi T^k_s}\right)^{3/2} \exp\left[-\frac{m_s \vpar^2}{2T_s^k}-\frac{\mu B}{T^k_s}\right].
\end{equation}
The source is uniform in the bi-normal coordinate $y$ but localized in the radial ($x$) and parallel ($z$) coordinates through $\eta^k_s$, so it is deliberately \emph{not} poloidally uniform. Given target input rates $\Gamma_{k,s}^\text{in}$ and powers $P_{k,s}^\text{in}$, the sub-source parameters are updated at each Runge--Kutta sub-step, denoted here $t_n$, to also compensate the instantaneous particle and power losses $\dot N^k_r(t_n)$ and $\dot E^k_r(t_n)$ of a (possibly different) species $r$ through a chosen set of boundaries,
\begin{align}
    \Gamma^k_s(t_{n+1}) &= \Gamma_{k,s}^\text{in} + \dot N^k_r(t_n), \\
    T^k_s(t_{n+1}) &= \frac{2}{3} \frac{P_{k,s}^\text{in} + \dot E^k_r(t_n)}{\Gamma^k_s(t_{n+1})}.
\end{align}
Allowing a source for species $s$ to compensate losses of species $r$ makes it possible to emulate ambipolar processes such as recycling.

In this work, two adaptive sub-sources per species are used. The \textit{core source} models heat transported from the tokamak core (not simulated here) and is localized near the inner radial boundary. It re-injects the particles lost through the inner boundary while adding a net power $P_\text{in}$ split evenly between electrons and ions. To mimic the spatial separation of the $\grad B$-drift-driven losses, the ion and electron core sources are offset in the parallel coordinate, at $z=\mp\pi/2$ respectively. The \textit{recycling source} emulates the ionization of neutrals generated by ions lost to the limiter and outer wall. It is localized on the high-field side, its particle rate is dynamically adjusted to compensate the ion losses through the outer radial boundary and the limiter, and it re-injects equal numbers of ions and electrons at a temperature of $10$~eV, contributing less than $1\%$ of the total power input.
The recycling source is a coarse model for plasma-neutral interactions. It reproduces the gross density injection at the high-field side, with a spatial distribution that is qualitatively consistent with kinetic neutral simulations \citep{Coroado2022ABoundary}, and uses a global recycling coefficient of unity.
It omits much of the physics that a self-consistent neutral model would capture. In particular, the ionization source of a fluid or kinetic neutral model depends nonlinearly on the local electron density and temperature and is non-local in the poloidal plane, since neutrals are ionized at a distance from where the recycled ions are born. The associated charge-exchange and ionization energy sinks provide a temperature-dependent cooling that is absent here. These effects play a leading-order role in the boundary dynamics of fluid edge codes, and their omission is a significant limitation of the present model. A coupled kinetic neutral model is under development and is expected to modify in particular the SOL temperature (Secs.~\ref{sec:results} and \ref{sec:conclusions}).

The adaptive source scheme could induce a feedback loop between the source and the losses, which could lead to oscillatory behavior. In practice, this is not observed in the simulations presented here. This likely indicates that the distance between each source and its related sinks is larger than the typical cross-field transport correlation length. The core sources are located near the inner radial boundary, on the side favoured by the radial $\grad B$ drift, while the recycling source is localized on the high-field side just inside the last closed flux surface, i.e. radially and poloidally separated from the outer wall and from the limiter whose losses it compensates.

The adaptivity acts only on the volume-integrated injection rates $\Gamma^k_s$ and $T^k_s$, and not on the local turbulent fluxes. The source remains a smooth, spatially localized function that does not track the instantaneous turbulent transport at the point where it is applied. The instantaneous compensation of the boundary losses therefore does not short-circuit the turbulent dynamics, which continues to set the profiles between the source and sink regions. Nevertheless, this remains a design choice that differs from a strictly constant-in-time forcing, and one must be careful that it does not artificially over-drive the system, i.e. push it further above the nonlinear turbulence threshold than a constant-power flux-driven simulation would. This concern is well documented in the comparison of gradient-driven and flux-driven gyrokinetic turbulence \citep{Rath2016,Peeters2016Gradient-drivenThreshold}, where the proximity to the nonlinear threshold controls both the mean transport and the level of mesoscale self-organization. In the present work, the net injected power $P_\text{in}$ and the target particle content are held fixed, and the adaptivity only redistributes them to compensate boundary losses on the sub-step timescale. The mean drive is therefore constant, but the fast compensation accelerates the approach to steady state and may bias the system toward stronger, more scale-separated turbulence than a slowly relaxing experiment. Quantifying this bias against a constant-source reference is left for future work.

\subsection{Boundary conditions}
\label{sec:appendix_bc}

The domain spans both closed and open field-line regions, which require different treatments. In the radial direction, the distribution function is set to zero outside the domain, so that particles drifting radially toward the core or the wall leave the domain at the inner and outer boundaries respectively, and no particles enter from the radial boundaries. For the potential, Dirichlet conditions ($\phi=0$~V) are applied at both radial boundaries. This removes the $\ExB$ drift at the boundary, so that radial particle losses there proceed solely through the $\grad B$ drift. 
While this can be expected at the outer radial boundary with a grounded wall, we note that this approach is not accurate for the inner radial boundary where the potential should be free to evolve up to the magnetic axis. Alternative boundary conditions, e.g. a Neumann condition or a radial-force-balance constraint on $E_r$, could be implemented in future work.

In the bi-normal direction, periodic boundary conditions are applied to both the distribution functions and the potential, following the standard flux-tube approach, which holds provided the bi-normal extent exceeds the turbulence correlation length.

In the parallel direction and closed flux surfaces, twist-and-shift boundary conditions connect the twisted ends of the flux tube \citep{Beer1995,Ball2020,Francisquez2024ConservativeConditions},
\begin{equation}
    F(x, y(z + L_z), z + L_z) = F(x, y(z), z),
\end{equation}
where $F$ is any field quantity and $L_z$ the parallel domain length. The conditions are applied at $z=\pm\pi$. In the open field-line region, conducting-sheath boundary conditions are applied to the potential, self-consistently building up a sheath potential that reflects low-energy electrons while allowing ions and high-energy electrons to leave the domain \citep{Shi_Hammett_Stoltzfus-Dueck_Hakim_2017}.

To conserve energy with the DG scheme, the potential must be continuous \citep{Francisquez2024ConservativeConditions}. The twist-and-shift condition is therefore applied to $\phi$ at $z=-\pi$ in the closed region so that it is twist-shift periodic with the upper end at $z=\pi$. Continuity must also be ensured at the limiter edge ($x=x_\text{LCFS}$, $z=\pm\pi$), where the twist-and-shift and sheath regions meet. This is achieved by imposing $\phi(x_\text{LCFS}, y, z=\pm\pi)=0$~V for all $y$. While the limiter corner could be treated more rigorously, this simplified approach is sufficient here: applying a different bias to the limiter-corner region does not strongly affect the bulk SOL turbulence.

\section{Ensemble statistics}
\label{sec:appendix_statistics}

We report in this appendix the quantities discussed in Secs.~\ref{sec:results} to \ref{sec:trapped_ions} for the whole ensemble, so that the trends described in the text can be read off directly and the dataset can be compared with other simulations or with experiments without reprocessing the raw output.
The full ensemble average and spread is given in Table~\ref{tab:ens_global}.
Both the mean and the median are listed there so that the reader can judge where the choice of statistic matters. For the profiles, the temperature ratios, the normalized gradients, the fluctuation levels and the electron transport channel, the two agree to within $4\%$, a small fraction of the standard deviation of each quantity. The profile distributions across the ensemble are close to symmetric and free of outliers. The mean and median measurements differ for the ion conductive channel, by $20\%$ for $\chi_i$ and $40\%$ for $q_i$, and to a lesser extent for the residual energy drift ($13\%$), the confinement time ($5\%$) and $\lambda_q$ ($3.5\%$), whose distributions have a tail towards large values. For these quantities the mean is pulled by a few simulations and the median is the statistic quoted in the text.
Tables~\ref{tab:ens_power}, \ref{tab:ens_delta} and \ref{tab:ens_kappa} present the marginal dependence on the injected power, the triangularity and the elongation, respectively, obtained by taking the median over the subset of simulations sharing a given parameter value. The standard deviation of the subset is also reported to indicate how much the quantity varies across the averaged parameters.

All quantities are computed with the conventions of Sec.~\ref{sec:steady_state}, i.e. profiles, gradients, fluxes and transport coefficients are evaluated at the outboard midplane ($z=0$) from the binormal and time average over the last $200~\mu$s of each simulation, at $r/a=0.9$ (edge), $r/a=1.0$ (separatrix) and $r/a=1.2$ (SOL).
The normalized gradients follow Eq.~\eqref{eq:norm_gradient}, the diffusivities Eq.~\eqref{eq:flux_gradient_relations} with the conductive heat flux of Eq.~\eqref{eq:conductive_heat_flux}, denoted $q_s$ in the tables, and the SOL heat-flux decay length $\lambda_q$ is the exponential fit described in Sec.~\ref{sec:sep_prof_grad}.
Two measures of the turbulence intensity are reported.
The relative fluctuation levels $\delta X_\text{rms}/\bar X$ labelled edge or SOL are the volume averages of Eq.~\eqref{eq:fluct_level} over $0.9<r/a<1$ and $1<r/a<1.2$, taken as the median of their time trace over the last $200\ \mu$s, while the one labelled $r/a=0.9$ is the binormal root-mean-square density fluctuation at that radius of the outboard midplane, averaged over the same window.
The energy confinement time $\tau_E = W/P_\text{loss}$ and the residual energy drift $\delta\dot W/P_\text{in}$ are defined in Sec.~\ref{sec:results}, $\Delta t$ is the mean time step and $t_\text{end}$ the final simulated time.

To make the comparison visible, the rows of Tables~\ref{tab:ens_power}, \ref{tab:ens_delta} and \ref{tab:ens_kappa} are shaded, from white for its smallest median to gray for its largest, when the tabulated parameter accounts for at least $10\%$ of the variance of the quantity over the ensemble.

\begin{table*}[ht]
    \centering
    \caption{Ensemble statistics over the 256 simulations: mean, standard deviation, median, and 10th and 90th percentiles. Definitions are given in the text.}
    \label{tab:ens_global}
    \begin{ruledtabular}
    \begin{tabular}{llccccc}
     & unit & mean & std & median & 10\% & 90\% \\
    \hline
    \multicolumn{7}{l}{\textit{Profiles at the outboard midplane}} \\
    $n_e^\text{sep}$ & $10^{18}$ m$^{-3}$ & 9.50 & 0.76 & 9.48 & 8.55 & 10.48 \\
    $T_e^\text{edge}$ & eV & 217 & 60 & 209 & 142 & 299 \\
    $T_e^\text{sep}$ & eV & 125 & 36 & 123 & 82 & 172 \\
    $T_e^\text{SOL}$ & eV & 56 & 19 & 54 & 34 & 84 \\
    $T_i^\text{edge}$ & eV & 207 & 58 & 207 & 127 & 291 \\
    $T_i^\text{sep}$ & eV & 170 & 48 & 169 & 106 & 233 \\
    $T_i^\text{SOL}$ & eV & 156 & 41 & 154 & 100 & 214 \\
    $T_i^\text{sep}/T_e^\text{sep}$ & -- & 1.38 & 0.25 & 1.35 & 1.10 & 1.77 \\
    $T_i^\text{edge}/T_i^\text{sep}$ & -- & 1.22 & 0.05 & 1.22 & 1.16 & 1.29 \\
    $T_i^\text{SOL}/T_i^\text{sep}$ & -- & 0.92 & 0.05 & 0.90 & 0.87 & 0.99 \\
    $T_{i,\perp}^\text{sep}/T_{i,\parallel}^\text{sep}$ & -- & 1.42 & 0.10 & 1.45 & 1.28 & 1.55 \\
    $T_{e,\perp}^\text{sep}/T_{e,\parallel}^\text{sep}$ & -- & 1.16 & 0.07 & 1.15 & 1.07 & 1.27 \\
    \hline
    \multicolumn{7}{l}{\textit{Normalized gradients, Eq.~\eqref{eq:norm_gradient}}} \\
    $R/L_{n_e}$ & -- & 25.5 & 1.3 & 25.6 & 23.8 & 27.1 \\
    $R/L_{T_e}$ & -- & 15.4 & 1.4 & 15.2 & 13.8 & 17.6 \\
    $R/L_{T_i}$ & -- & 6.4 & 1.3 & 6.5 & 4.9 & 8.1 \\
    \hline
    \multicolumn{7}{l}{\textit{Fluctuation levels}} \\
    $\delta n_{e,\text{rms}}/\bar n_e|_\text{edge}$ & -- & 0.152 & 0.010 & 0.153 & 0.140 & 0.164 \\
    $\delta n_{e,\text{rms}}/\bar n_e|_\text{SOL}$ & -- & 0.255 & 0.030 & 0.250 & 0.219 & 0.296 \\
    $\delta T_{e,\text{rms}}/\bar T_e|_\text{edge}$ & -- & 0.041 & 0.007 & 0.040 & 0.033 & 0.050 \\
    $\delta T_{e,\text{rms}}/\bar T_e|_\text{SOL}$ & -- & 0.205 & 0.021 & 0.206 & 0.181 & 0.230 \\
    $\delta T_{i,\text{rms}}/\bar T_i|_\text{edge}$ & -- & 0.045 & 0.006 & 0.045 & 0.036 & 0.053 \\
    $\delta T_{i,\text{rms}}/\bar T_i|_\text{SOL}$ & -- & 0.089 & 0.021 & 0.084 & 0.064 & 0.120 \\
    $\delta n_{e,\text{rms}}/\bar n_e|_{r/a=0.9}$ & -- & 0.056 & 0.010 & 0.056 & 0.042 & 0.070 \\
    \hline
    \multicolumn{7}{l}{\textit{Separatrix transport}} \\
    $\Gamma_e^\text{sep}$ & $10^{20}$ m$^{-2}$s$^{-1}$ & 10.47 & 1.49 & 10.51 & 8.60 & 12.27 \\
    $q_e^\text{sep}$ & kW m$^{-2}$ & 23.3 & 4.9 & 23.5 & 16.7 & 29.2 \\
    $q_i^\text{sep}$ & kW m$^{-2}$ & -1.7 & 2.4 & -1.2 & -5.3 & 0.9 \\
    $D_e$ & m$^2$/s & 1.12 & 0.22 & 1.11 & 0.83 & 1.38 \\
    $\chi_e$ & m$^2$/s & 4.20 & 0.92 & 4.24 & 2.94 & 5.34 \\
    $D_i$ & m$^2$/s & 0.80 & 0.38 & 0.82 & 0.33 & 1.27 \\
    $\chi_i$ & m$^2$/s & -0.70 & 0.97 & -0.58 & -2.08 & 0.43 \\
    $q_e/(\tfrac{3}{2}\overline{T}_e\Gamma_e)$ & -- & 0.77 & 0.16 & 0.77 & 0.57 & 0.95 \\
    $\lambda_q$ & mm & 4.12 & 1.03 & 3.98 & 2.99 & 5.58 \\
    \hline
    \multicolumn{7}{l}{\textit{Global quantities}} \\
    $\tau_E$ & ms & 0.90 & 0.18 & 0.86 & 0.71 & 1.17 \\
    $\delta\dot W/P_\text{in}$ & -- & 0.078 & 0.040 & 0.069 & 0.037 & 0.132 \\
    $\Delta t$ & ns & 3.18 & 0.51 & 3.13 & 2.57 & 3.96 \\
    $t_\text{end}$ & ms & 1.64 & 0.28 & 1.63 & 1.23 & 1.99 \\
    \end{tabular}
    \end{ruledtabular}
\end{table*}

\begin{table*}[ht]
    \centering
    \caption{Medians over the 64 shaping configurations at each injected power, with the standard deviation of the subset in percent of the median. Rows are shaded, linearly from white for the smallest value to gray for the largest, when the injected power accounts for at least 10\% of the variance of the quantity over the ensemble.}
    \label{tab:ens_power}
    \begin{tabularx}{\textwidth}{ll*{4}{>{\centering\arraybackslash}X}}
    \hline\hline
     & unit & $P_\text{in}=0.1$ MW & $P_\text{in}=0.5$ MW & $P_\text{in}=1.0$ MW & $P_\text{in}=5.0$ MW \\
    \hline
    \multicolumn{6}{l}{\textit{Profiles at the outboard midplane}} \\
    $n_e^\text{sep}$ & $10^{18}$ m$^{-3}$ & \cellcolor[gray]{0.70}$9.74 \pm 6.8\%$ & \cellcolor[gray]{0.80}$9.53 \pm 6.1\%$ & \cellcolor[gray]{0.89}$9.34 \pm 7.3\%$ & \cellcolor[gray]{1.00}$9.10 \pm 9.8\%$ \\
    $T_e^\text{edge}$ & eV & \cellcolor[gray]{1.00}$152 \pm 22\%$ & \cellcolor[gray]{0.89}$202 \pm 12\%$ & \cellcolor[gray]{0.83}$228 \pm 12\%$ & \cellcolor[gray]{0.70}$289 \pm 14\%$ \\
    $T_e^\text{sep}$ & eV & \cellcolor[gray]{1.00}$91 \pm 28\%$ & \cellcolor[gray]{0.90}$117 \pm 17\%$ & \cellcolor[gray]{0.84}$132 \pm 14\%$ & \cellcolor[gray]{0.70}$166 \pm 14\%$ \\
    $T_e^\text{SOL}$ & eV & \cellcolor[gray]{1.00}$37 \pm 27\%$ & \cellcolor[gray]{0.90}$51 \pm 18\%$ & \cellcolor[gray]{0.84}$59 \pm 16\%$ & \cellcolor[gray]{0.70}$77 \pm 19\%$ \\
    $T_i^\text{edge}$ & eV & \cellcolor[gray]{1.00}$128 \pm 3.6\%$ & \cellcolor[gray]{0.87}$191 \pm 6.0\%$ & \cellcolor[gray]{0.82}$220 \pm 7.7\%$ & \cellcolor[gray]{0.70}$278 \pm 11\%$ \\
    $T_i^\text{sep}$ & eV & \cellcolor[gray]{1.00}$108 \pm 7.2\%$ & \cellcolor[gray]{0.87}$159 \pm 7.8\%$ & \cellcolor[gray]{0.82}$180 \pm 9.7\%$ & \cellcolor[gray]{0.70}$224 \pm 15\%$ \\
    $T_i^\text{SOL}$ & eV & \cellcolor[gray]{1.00}$104 \pm 15\%$ & \cellcolor[gray]{0.87}$145 \pm 9.9\%$ & \cellcolor[gray]{0.82}$162 \pm 11\%$ & \cellcolor[gray]{0.70}$203 \pm 15\%$ \\
    $T_i^\text{sep}/T_e^\text{sep}$ & -- & $1.17 \pm 30\%$ & $1.36 \pm 17\%$ & $1.41 \pm 13\%$ & $1.33 \pm 11\%$ \\
    $T_i^\text{edge}/T_i^\text{sep}$ & -- & \cellcolor[gray]{1.00}$1.19 \pm 5.0\%$ & \cellcolor[gray]{0.90}$1.21 \pm 3.4\%$ & \cellcolor[gray]{0.86}$1.22 \pm 3.1\%$ & \cellcolor[gray]{0.70}$1.25 \pm 4.1\%$ \\
    $T_i^\text{SOL}/T_i^\text{sep}$ & -- & \cellcolor[gray]{0.70}$0.96 \pm 7.8\%$ & \cellcolor[gray]{0.95}$0.90 \pm 4.1\%$ & \cellcolor[gray]{0.96}$0.90 \pm 3.1\%$ & \cellcolor[gray]{1.00}$0.89 \pm 3.0\%$ \\
    $T_{i,\perp}^\text{sep}/T_{i,\parallel}^\text{sep}$ & -- & \cellcolor[gray]{0.70}$1.49 \pm 3.9\%$ & \cellcolor[gray]{0.74}$1.47 \pm 5.4\%$ & \cellcolor[gray]{0.78}$1.44 \pm 5.6\%$ & \cellcolor[gray]{1.00}$1.31 \pm 6.0\%$ \\
    $T_{e,\perp}^\text{sep}/T_{e,\parallel}^\text{sep}$ & -- & \cellcolor[gray]{1.00}$1.09 \pm 3.4\%$ & \cellcolor[gray]{0.92}$1.13 \pm 2.5\%$ & \cellcolor[gray]{0.86}$1.16 \pm 2.4\%$ & \cellcolor[gray]{0.70}$1.25 \pm 4.2\%$ \\
    \hline
    \multicolumn{6}{l}{\textit{Normalized gradients, Eq.~\eqref{eq:norm_gradient}}} \\
    $R/L_{n_e}$ & -- & $25.6 \pm 3.6\%$ & $25.5 \pm 4.6\%$ & $25.5 \pm 5.3\%$ & $25.9 \pm 5.8\%$ \\
    $R/L_{T_e}$ & -- & $14.9 \pm 12\%$ & $15.4 \pm 9.7\%$ & $15.5 \pm 8.4\%$ & $15.2 \pm 6.8\%$ \\
    $R/L_{T_i}$ & -- & \cellcolor[gray]{1.00}$5.8 \pm 26\%$ & \cellcolor[gray]{0.90}$6.3 \pm 16\%$ & \cellcolor[gray]{0.86}$6.5 \pm 14\%$ & \cellcolor[gray]{0.70}$7.3 \pm 17\%$ \\
    \hline
    \multicolumn{6}{l}{\textit{Fluctuation levels}} \\
    $\delta n_{e,\text{rms}}/\bar n_e|_\text{edge}$ & -- & \cellcolor[gray]{0.73}$0.155 \pm 5.3\%$ & \cellcolor[gray]{0.70}$0.156 \pm 5.5\%$ & \cellcolor[gray]{0.75}$0.154 \pm 5.2\%$ & \cellcolor[gray]{1.00}$0.144 \pm 7.9\%$ \\
    $\delta n_{e,\text{rms}}/\bar n_e|_\text{SOL}$ & -- & $0.240 \pm 11\%$ & $0.247 \pm 11\%$ & $0.249 \pm 12\%$ & $0.266 \pm 11\%$ \\
    $\delta T_{e,\text{rms}}/\bar T_e|_\text{edge}$ & -- & $0.040 \pm 20\%$ & $0.042 \pm 15\%$ & $0.040 \pm 15\%$ & $0.038 \pm 16\%$ \\
    $\delta T_{e,\text{rms}}/\bar T_e|_\text{SOL}$ & -- & \cellcolor[gray]{0.97}$0.199 \pm 10\%$ & \cellcolor[gray]{0.73}$0.212 \pm 7.1\%$ & \cellcolor[gray]{0.70}$0.213 \pm 9.4\%$ & \cellcolor[gray]{1.00}$0.197 \pm 11\%$ \\
    $\delta T_{i,\text{rms}}/\bar T_i|_\text{edge}$ & -- & \cellcolor[gray]{1.00}$0.037 \pm 7.3\%$ & \cellcolor[gray]{0.85}$0.044 \pm 7.9\%$ & \cellcolor[gray]{0.79}$0.047 \pm 8.3\%$ & \cellcolor[gray]{0.70}$0.051 \pm 6.4\%$ \\
    $\delta T_{i,\text{rms}}/\bar T_i|_\text{SOL}$ & -- & \cellcolor[gray]{1.00}$0.066 \pm 12\%$ & \cellcolor[gray]{0.90}$0.081 \pm 14\%$ & \cellcolor[gray]{0.83}$0.091 \pm 16\%$ & \cellcolor[gray]{0.70}$0.109 \pm 18\%$ \\
    $\delta n_{e,\text{rms}}/\bar n_e|_{r/a=0.9}$ & -- & \cellcolor[gray]{0.76}$0.057 \pm 16\%$ & \cellcolor[gray]{0.70}$0.058 \pm 15\%$ & \cellcolor[gray]{0.73}$0.057 \pm 18\%$ & \cellcolor[gray]{1.00}$0.051 \pm 22\%$ \\
    \hline
    \multicolumn{6}{l}{\textit{Separatrix transport}} \\
    $\Gamma_e^\text{sep}$ & $10^{20}$ m$^{-2}$s$^{-1}$ & \cellcolor[gray]{0.90}$10.27 \pm 14\%$ & \cellcolor[gray]{0.70}$11.27 \pm 11\%$ & \cellcolor[gray]{0.83}$10.61 \pm 14\%$ & \cellcolor[gray]{1.00}$9.73 \pm 16\%$ \\
    $q_e^\text{sep}$ & kW m$^{-2}$ & \cellcolor[gray]{1.00}$19.7 \pm 27\%$ & \cellcolor[gray]{0.70}$25.5 \pm 16\%$ & \cellcolor[gray]{0.73}$24.9 \pm 15\%$ & \cellcolor[gray]{0.80}$23.5 \pm 13\%$ \\
    $q_i^\text{sep}$ & kW m$^{-2}$ & \cellcolor[gray]{0.70}$0.0 \pm 1.1$ & \cellcolor[gray]{0.77}$-0.8 \pm 1.2$ & \cellcolor[gray]{0.87}$-2.1 \pm 84\%$ & \cellcolor[gray]{1.00}$-3.6 \pm 76\%$ \\
    $D_e$ & m$^2$/s & $1.05 \pm 19\%$ & $1.20 \pm 16\%$ & $1.13 \pm 20\%$ & $1.05 \pm 22\%$ \\
    $\chi_e$ & m$^2$/s & \cellcolor[gray]{0.70}$4.69 \pm 20\%$ & \cellcolor[gray]{0.70}$4.67 \pm 13\%$ & \cellcolor[gray]{0.79}$4.27 \pm 14\%$ & \cellcolor[gray]{1.00}$3.28 \pm 18\%$ \\
    $D_i$ & m$^2$/s & $0.84 \pm 28\%$ & $0.94 \pm 34\%$ & $0.83 \pm 46\%$ & $0.65 \pm 73\%$ \\
    $\chi_i$ & m$^2$/s & \cellcolor[gray]{0.70}$0.00 \pm 0.92$ & \cellcolor[gray]{0.81}$-0.39 \pm 0.61$ & \cellcolor[gray]{0.93}$-0.86 \pm 83\%$ & \cellcolor[gray]{1.00}$-1.10 \pm 99\%$ \\
    $q_e/(\tfrac{3}{2}\overline{T}_e\Gamma_e)$ & -- & \cellcolor[gray]{0.70}$0.89 \pm 11\%$ & \cellcolor[gray]{0.76}$0.82 \pm 18\%$ & \cellcolor[gray]{0.82}$0.77 \pm 15\%$ & \cellcolor[gray]{1.00}$0.59 \pm 20\%$ \\
    $\lambda_q$ & mm & $3.92 \pm 22\%$ & $4.19 \pm 27\%$ & $4.01 \pm 25\%$ & $3.58 \pm 29\%$ \\
    \hline
    \multicolumn{6}{l}{\textit{Global quantities}} \\
    $\tau_E$ & ms & $0.93 \pm 17\%$ & $0.83 \pm 19\%$ & $0.84 \pm 22\%$ & $0.84 \pm 23\%$ \\
    $\delta\dot W/P_\text{in}$ & -- & \cellcolor[gray]{0.70}$0.103 \pm 41\%$ & \cellcolor[gray]{0.94}$0.065 \pm 59\%$ & \cellcolor[gray]{0.96}$0.060 \pm 52\%$ & \cellcolor[gray]{1.00}$0.055 \pm 42\%$ \\
    $\Delta t$ & ns & $3.14 \pm 15\%$ & $3.12 \pm 17\%$ & $3.14 \pm 17\%$ & $3.11 \pm 16\%$ \\
    $t_\text{end}$ & ms & $1.63 \pm 18\%$ & $1.62 \pm 17\%$ & $1.62 \pm 17\%$ & $1.66 \pm 17\%$ \\
    \hline\hline
    \end{tabularx}
\end{table*}

\begin{table*}[ht]
    \centering
    \caption{Medians over the 32 simulations at each triangularity $\delta$, with the standard deviation of the subset in percent of the median. Rows are shaded, linearly from white for the smallest value to gray for the largest, when $\delta$ accounts for at least 10\% of the variance of the quantity over the ensemble. Units as in Table~\ref{tab:ens_global}.}
    \label{tab:ens_delta}
    \footnotesize
    \begin{tabularx}{\textwidth}{l*{8}{>{\centering\arraybackslash}X}}
    \hline\hline
     & $-0.60$ & $-0.45$ & $-0.30$ & $-0.15$ & $+0.15$ & $+0.30$ & $+0.45$ & $+0.60$ \\
    \hline
    \multicolumn{9}{l}{\textit{Profiles at the outboard midplane}} \\
    $n_e^\text{sep}$ & $9.61 \pm 6.8\%$ & $9.69 \pm 6.4\%$ & $9.45 \pm 6.2\%$ & $9.49 \pm 5.4\%$ & $9.36 \pm 7.5\%$ & $9.15 \pm 8.8\%$ & $9.35 \pm 10\%$ & $9.51 \pm 10\%$ \\
    $T_e^\text{edge}$ & $192 \pm 24\%$ & $200 \pm 23\%$ & $206 \pm 24\%$ & $206 \pm 27\%$ & $216 \pm 29\%$ & $206 \pm 32\%$ & $232 \pm 29\%$ & $241 \pm 28\%$ \\
    $T_e^\text{sep}$ & $113 \pm 24\%$ & $124 \pm 23\%$ & $125 \pm 26\%$ & $122 \pm 30\%$ & $128 \pm 30\%$ & $109 \pm 36\%$ & $131 \pm 29\%$ & $140 \pm 27\%$ \\
    $T_e^\text{SOL}$ & $55 \pm 23\%$ & $56 \pm 26\%$ & $57 \pm 28\%$ & $53 \pm 35\%$ & $52 \pm 37\%$ & $44 \pm 45\%$ & $55 \pm 40\%$ & $54 \pm 44\%$ \\
    $T_i^\text{edge}$ & $193 \pm 23\%$ & $196 \pm 23\%$ & $198 \pm 24\%$ & $201 \pm 25\%$ & $213 \pm 28\%$ & $218 \pm 30\%$ & $219 \pm 31\%$ & $229 \pm 31\%$ \\
    $T_i^\text{sep}$ & \cellcolor[gray]{1.00}$156 \pm 21\%$ & \cellcolor[gray]{0.99}$156 \pm 22\%$ & \cellcolor[gray]{0.97}$159 \pm 23\%$ & \cellcolor[gray]{0.95}$162 \pm 24\%$ & \cellcolor[gray]{0.86}$173 \pm 26\%$ & \cellcolor[gray]{0.78}$182 \pm 30\%$ & \cellcolor[gray]{0.76}$185 \pm 31\%$ & \cellcolor[gray]{0.70}$192 \pm 31\%$ \\
    $T_i^\text{SOL}$ & \cellcolor[gray]{1.00}$141 \pm 20\%$ & \cellcolor[gray]{0.99}$142 \pm 21\%$ & \cellcolor[gray]{0.96}$146 \pm 21\%$ & \cellcolor[gray]{0.94}$148 \pm 22\%$ & \cellcolor[gray]{0.87}$157 \pm 25\%$ & \cellcolor[gray]{0.80}$165 \pm 28\%$ & \cellcolor[gray]{0.75}$171 \pm 27\%$ & \cellcolor[gray]{0.70}$178 \pm 28\%$ \\
    $T_i^\text{sep}/T_e^\text{sep}$ & \cellcolor[gray]{0.95}$1.33 \pm 12\%$ & \cellcolor[gray]{0.98}$1.29 \pm 15\%$ & \cellcolor[gray]{1.00}$1.26 \pm 21\%$ & \cellcolor[gray]{0.99}$1.28 \pm 24\%$ & \cellcolor[gray]{0.94}$1.36 \pm 19\%$ & \cellcolor[gray]{0.70}$1.73 \pm 16\%$ & \cellcolor[gray]{0.90}$1.42 \pm 11\%$ & \cellcolor[gray]{0.92}$1.39 \pm 11\%$ \\
    $T_i^\text{edge}/T_i^\text{sep}$ & \cellcolor[gray]{0.72}$1.25 \pm 4.5\%$ & \cellcolor[gray]{0.70}$1.25 \pm 4.3\%$ & \cellcolor[gray]{0.70}$1.25 \pm 4.0\%$ & \cellcolor[gray]{0.70}$1.25 \pm 3.9\%$ & \cellcolor[gray]{0.78}$1.23 \pm 3.4\%$ & \cellcolor[gray]{0.97}$1.19 \pm 3.2\%$ & \cellcolor[gray]{0.96}$1.20 \pm 2.9\%$ & \cellcolor[gray]{1.00}$1.19 \pm 2.6\%$ \\
    $T_i^\text{SOL}/T_i^\text{sep}$ & $0.90 \pm 2.6\%$ & $0.90 \pm 3.1\%$ & $0.90 \pm 3.6\%$ & $0.90 \pm 4.3\%$ & $0.91 \pm 5.9\%$ & $0.92 \pm 6.1\%$ & $0.91 \pm 8.2\%$ & $0.92 \pm 9.5\%$ \\
    $T_{i,\perp}^\text{sep}/T_{i,\parallel}^\text{sep}$ & \cellcolor[gray]{0.75}$1.47 \pm 7.7\%$ & \cellcolor[gray]{0.79}$1.45 \pm 5.9\%$ & \cellcolor[gray]{0.78}$1.45 \pm 5.3\%$ & \cellcolor[gray]{0.77}$1.46 \pm 4.9\%$ & \cellcolor[gray]{0.76}$1.46 \pm 5.9\%$ & \cellcolor[gray]{0.70}$1.50 \pm 7.3\%$ & \cellcolor[gray]{0.88}$1.39 \pm 7.1\%$ & \cellcolor[gray]{1.00}$1.32 \pm 7.4\%$ \\
    $T_{e,\perp}^\text{sep}/T_{e,\parallel}^\text{sep}$ & $1.13 \pm 4.7\%$ & $1.13 \pm 4.8\%$ & $1.15 \pm 5.0\%$ & $1.14 \pm 5.7\%$ & $1.16 \pm 6.7\%$ & $1.14 \pm 7.1\%$ & $1.16 \pm 6.9\%$ & $1.18 \pm 7.4\%$ \\
    \hline
    \multicolumn{9}{l}{\textit{Normalized gradients, Eq.~\eqref{eq:norm_gradient}}} \\
    $R/L_{n_e}$ & \cellcolor[gray]{1.00}$24.8 \pm 4.6\%$ & \cellcolor[gray]{0.95}$25.0 \pm 4.5\%$ & \cellcolor[gray]{0.92}$25.3 \pm 4.1\%$ & \cellcolor[gray]{0.88}$25.5 \pm 3.5\%$ & \cellcolor[gray]{0.82}$25.9 \pm 4.0\%$ & \cellcolor[gray]{0.70}$26.6 \pm 4.0\%$ & \cellcolor[gray]{0.77}$26.1 \pm 5.0\%$ & \cellcolor[gray]{0.79}$26.0 \pm 5.2\%$ \\
    $R/L_{T_e}$ & \cellcolor[gray]{0.95}$14.8 \pm 5.5\%$ & \cellcolor[gray]{0.99}$14.4 \pm 7.2\%$ & \cellcolor[gray]{1.00}$14.3 \pm 10.0\%$ & \cellcolor[gray]{0.95}$14.8 \pm 12\%$ & \cellcolor[gray]{0.87}$15.7 \pm 8.5\%$ & \cellcolor[gray]{0.70}$17.6 \pm 8.1\%$ & \cellcolor[gray]{0.87}$15.6 \pm 4.0\%$ & \cellcolor[gray]{0.90}$15.4 \pm 4.0\%$ \\
    $R/L_{T_i}$ & \cellcolor[gray]{0.72}$7.1 \pm 19\%$ & \cellcolor[gray]{0.70}$7.3 \pm 18\%$ & \cellcolor[gray]{0.70}$7.3 \pm 16\%$ & \cellcolor[gray]{0.70}$7.2 \pm 17\%$ & \cellcolor[gray]{0.78}$6.9 \pm 16\%$ & \cellcolor[gray]{0.97}$5.8 \pm 17\%$ & \cellcolor[gray]{0.95}$5.9 \pm 15\%$ & \cellcolor[gray]{1.00}$5.7 \pm 14\%$ \\
    \hline
    \multicolumn{9}{l}{\textit{Fluctuation levels}} \\
    $\delta n_{e,\text{rms}}/\bar n_e|_\text{edge}$ & \cellcolor[gray]{0.75}$0.154 \pm 6.1\%$ & \cellcolor[gray]{0.71}$0.156 \pm 5.2\%$ & \cellcolor[gray]{0.73}$0.155 \pm 5.3\%$ & \cellcolor[gray]{0.70}$0.156 \pm 6.1\%$ & \cellcolor[gray]{0.75}$0.154 \pm 7.0\%$ & \cellcolor[gray]{0.89}$0.150 \pm 6.1\%$ & \cellcolor[gray]{0.90}$0.149 \pm 6.8\%$ & \cellcolor[gray]{1.00}$0.146 \pm 7.6\%$ \\
    $\delta n_{e,\text{rms}}/\bar n_e|_\text{SOL}$ & \cellcolor[gray]{1.00}$0.224 \pm 6.5\%$ & \cellcolor[gray]{0.99}$0.227 \pm 4.9\%$ & \cellcolor[gray]{0.97}$0.232 \pm 5.8\%$ & \cellcolor[gray]{0.93}$0.240 \pm 7.1\%$ & \cellcolor[gray]{0.86}$0.259 \pm 7.0\%$ & \cellcolor[gray]{0.80}$0.273 \pm 6.4\%$ & \cellcolor[gray]{0.76}$0.282 \pm 5.8\%$ & \cellcolor[gray]{0.70}$0.296 \pm 6.0\%$ \\
    $\delta T_{e,\text{rms}}/\bar T_e|_\text{edge}$ & \cellcolor[gray]{0.70}$0.045 \pm 12\%$ & \cellcolor[gray]{0.87}$0.040 \pm 15\%$ & \cellcolor[gray]{0.97}$0.038 \pm 22\%$ & \cellcolor[gray]{0.86}$0.041 \pm 20\%$ & \cellcolor[gray]{0.87}$0.040 \pm 18\%$ & \cellcolor[gray]{0.73}$0.044 \pm 14\%$ & \cellcolor[gray]{0.96}$0.038 \pm 12\%$ & \cellcolor[gray]{1.00}$0.037 \pm 12\%$ \\
    $\delta T_{e,\text{rms}}/\bar T_e|_\text{SOL}$ & \cellcolor[gray]{1.00}$0.193 \pm 9.1\%$ & \cellcolor[gray]{0.98}$0.195 \pm 6.1\%$ & \cellcolor[gray]{0.96}$0.197 \pm 6.0\%$ & \cellcolor[gray]{0.84}$0.207 \pm 9.2\%$ & \cellcolor[gray]{0.74}$0.216 \pm 9.0\%$ & \cellcolor[gray]{0.70}$0.219 \pm 9.6\%$ & \cellcolor[gray]{0.75}$0.215 \pm 11\%$ & \cellcolor[gray]{0.77}$0.213 \pm 12\%$ \\
    $\delta T_{i,\text{rms}}/\bar T_i|_\text{edge}$ & $0.045 \pm 14\%$ & $0.045 \pm 14\%$ & $0.046 \pm 13\%$ & $0.044 \pm 14\%$ & $0.044 \pm 14\%$ & $0.045 \pm 13\%$ & $0.045 \pm 14\%$ & $0.045 \pm 12\%$ \\
    $\delta T_{i,\text{rms}}/\bar T_i|_\text{SOL}$ & \cellcolor[gray]{1.00}$0.079 \pm 20\%$ & \cellcolor[gray]{1.00}$0.079 \pm 21\%$ & \cellcolor[gray]{1.00}$0.079 \pm 21\%$ & \cellcolor[gray]{1.00}$0.079 \pm 22\%$ & \cellcolor[gray]{0.92}$0.086 \pm 23\%$ & \cellcolor[gray]{0.86}$0.090 \pm 24\%$ & \cellcolor[gray]{0.82}$0.094 \pm 24\%$ & \cellcolor[gray]{0.70}$0.103 \pm 23\%$ \\
    $\delta n_{e,\text{rms}}/\bar n_e|_{r/a=0.9}$ & \cellcolor[gray]{0.70}$0.066 \pm 8.2\%$ & \cellcolor[gray]{0.71}$0.065 \pm 8.2\%$ & \cellcolor[gray]{0.74}$0.063 \pm 10\%$ & \cellcolor[gray]{0.79}$0.060 \pm 13\%$ & \cellcolor[gray]{0.89}$0.053 \pm 15\%$ & \cellcolor[gray]{1.00}$0.047 \pm 16\%$ & \cellcolor[gray]{0.95}$0.050 \pm 15\%$ & \cellcolor[gray]{0.95}$0.050 \pm 16\%$ \\
    \hline
    \multicolumn{9}{l}{\textit{Separatrix transport}} \\
    $\Gamma_e^\text{sep}$ & \cellcolor[gray]{0.79}$10.75 \pm 14\%$ & \cellcolor[gray]{0.70}$11.25 \pm 13\%$ & \cellcolor[gray]{0.81}$10.62 \pm 15\%$ & \cellcolor[gray]{0.93}$9.97 \pm 14\%$ & \cellcolor[gray]{0.86}$10.33 \pm 16\%$ & \cellcolor[gray]{1.00}$9.56 \pm 11\%$ & \cellcolor[gray]{0.84}$10.47 \pm 12\%$ & \cellcolor[gray]{0.75}$10.99 \pm 12\%$ \\
    $q_e^\text{sep}$ & \cellcolor[gray]{0.99}$21.8 \pm 19\%$ & \cellcolor[gray]{0.94}$22.9 \pm 16\%$ & \cellcolor[gray]{0.92}$23.5 \pm 19\%$ & \cellcolor[gray]{0.89}$24.0 \pm 23\%$ & \cellcolor[gray]{0.92}$23.4 \pm 25\%$ & \cellcolor[gray]{1.00}$21.6 \pm 20\%$ & \cellcolor[gray]{0.83}$25.6 \pm 17\%$ & \cellcolor[gray]{0.70}$28.5 \pm 15\%$ \\
    $q_i^\text{sep}$ & \cellcolor[gray]{0.76}$-1.0 \pm 1.4$ & \cellcolor[gray]{0.70}$-0.5 \pm 1.5$ & \cellcolor[gray]{0.74}$-0.9 \pm 1.1$ & \cellcolor[gray]{0.74}$-0.9 \pm 1.4$ & \cellcolor[gray]{0.82}$-1.5 \pm 1.9$ & \cellcolor[gray]{1.00}$-3.1 \pm 79\%$ & \cellcolor[gray]{0.89}$-2.1 \pm 3.0$ & \cellcolor[gray]{0.91}$-2.4 \pm 3.3$ \\
    $D_e$ & \cellcolor[gray]{0.70}$1.22 \pm 18\%$ & \cellcolor[gray]{0.70}$1.23 \pm 19\%$ & \cellcolor[gray]{0.79}$1.13 \pm 18\%$ & \cellcolor[gray]{0.80}$1.13 \pm 14\%$ & \cellcolor[gray]{0.82}$1.09 \pm 19\%$ & \cellcolor[gray]{1.00}$0.91 \pm 16\%$ & \cellcolor[gray]{0.87}$1.04 \pm 18\%$ & \cellcolor[gray]{0.81}$1.11 \pm 18\%$ \\
    $\chi_e$ & $4.30 \pm 14\%$ & $4.39 \pm 18\%$ & $4.41 \pm 19\%$ & $4.27 \pm 22\%$ & $3.95 \pm 23\%$ & $3.50 \pm 24\%$ & $4.33 \pm 23\%$ & $4.39 \pm 25\%$ \\
    $D_i$ & \cellcolor[gray]{0.70}$1.20 \pm 18\%$ & \cellcolor[gray]{0.71}$1.17 \pm 20\%$ & \cellcolor[gray]{0.77}$1.02 \pm 20\%$ & \cellcolor[gray]{0.82}$0.91 \pm 16\%$ & \cellcolor[gray]{0.89}$0.72 \pm 27\%$ & \cellcolor[gray]{0.95}$0.56 \pm 35\%$ & \cellcolor[gray]{0.96}$0.55 \pm 47\%$ & \cellcolor[gray]{1.00}$0.44 \pm 72\%$ \\
    $\chi_i$ & \cellcolor[gray]{0.76}$-0.42 \pm 0.74$ & \cellcolor[gray]{0.70}$-0.19 \pm 0.69$ & \cellcolor[gray]{0.74}$-0.34 \pm 0.52$ & \cellcolor[gray]{0.75}$-0.39 \pm 0.56$ & \cellcolor[gray]{0.83}$-0.71 \pm 0.73$ & \cellcolor[gray]{1.00}$-1.40 \pm 65\%$ & \cellcolor[gray]{0.93}$-1.11 \pm 1.24$ & \cellcolor[gray]{0.97}$-1.27 \pm 1.37$ \\
    $q_e/(\tfrac{3}{2}\overline{T}_e\Gamma_e)$ & $0.73 \pm 14\%$ & $0.73 \pm 16\%$ & $0.72 \pm 23\%$ & $0.79 \pm 25\%$ & $0.81 \pm 18\%$ & $0.88 \pm 17\%$ & $0.82 \pm 21\%$ & $0.81 \pm 21\%$ \\
    $\lambda_q$ & $4.07 \pm 24\%$ & $3.82 \pm 21\%$ & $4.02 \pm 26\%$ & $3.87 \pm 25\%$ & $3.83 \pm 26\%$ & $4.56 \pm 27\%$ & $3.77 \pm 28\%$ & $3.80 \pm 24\%$ \\
    \hline
    \multicolumn{9}{l}{\textit{Global quantities}} \\
    $\tau_E$ & \cellcolor[gray]{0.98}$0.81 \pm 11\%$ & \cellcolor[gray]{1.00}$0.79 \pm 12\%$ & \cellcolor[gray]{0.99}$0.80 \pm 11\%$ & \cellcolor[gray]{0.97}$0.82 \pm 12\%$ & \cellcolor[gray]{0.89}$0.92 \pm 15\%$ & \cellcolor[gray]{0.70}$1.13 \pm 16\%$ & \cellcolor[gray]{0.85}$0.96 \pm 19\%$ & \cellcolor[gray]{0.85}$0.96 \pm 21\%$ \\
    $\delta\dot W/P_\text{in}$ & $0.069 \pm 43\%$ & $0.075 \pm 49\%$ & $0.066 \pm 76\%$ & $0.073 \pm 62\%$ & $0.087 \pm 47\%$ & $0.056 \pm 67\%$ & $0.059 \pm 63\%$ & $0.062 \pm 50\%$ \\
    $\Delta t$ & \cellcolor[gray]{0.87}$3.04 \pm 12\%$ & \cellcolor[gray]{0.80}$3.28 \pm 13\%$ & \cellcolor[gray]{0.74}$3.49 \pm 14\%$ & \cellcolor[gray]{0.70}$3.62 \pm 14\%$ & \cellcolor[gray]{0.75}$3.44 \pm 13\%$ & \cellcolor[gray]{0.84}$3.12 \pm 6.1\%$ & \cellcolor[gray]{0.93}$2.83 \pm 10\%$ & \cellcolor[gray]{1.00}$2.57 \pm 8.0\%$ \\
    $t_\text{end}$ & \cellcolor[gray]{0.87}$1.54 \pm 16\%$ & \cellcolor[gray]{0.81}$1.68 \pm 15\%$ & \cellcolor[gray]{0.75}$1.83 \pm 12\%$ & \cellcolor[gray]{0.72}$1.89 \pm 11\%$ & \cellcolor[gray]{0.76}$1.81 \pm 11\%$ & \cellcolor[gray]{0.70}$1.95 \pm 10\%$ & \cellcolor[gray]{0.91}$1.44 \pm 15\%$ & \cellcolor[gray]{1.00}$1.21 \pm 14\%$ \\
    \hline\hline
    \end{tabularx}
\end{table*}

\begin{table*}[ht]
    \centering
    \caption{Medians over the 32 simulations at each elongation $\kappa$, with the standard deviation of the subset in percent of the median. Rows are shaded, linearly from white for the smallest value to gray for the largest, when $\kappa$ accounts for at least 10\% of the variance of the quantity over the ensemble. Units as in Table~\ref{tab:ens_global}.}
    \label{tab:ens_kappa}
    \footnotesize
    \begin{tabularx}{\textwidth}{l*{8}{>{\centering\arraybackslash}X}}
    \hline\hline
     & $1.1$ & $1.2$ & $1.3$ & $1.4$ & $1.5$ & $1.6$ & $1.7$ & $1.8$ \\
    \hline
    \multicolumn{9}{l}{\textit{Profiles at the outboard midplane}} \\
    $n_e^\text{sep}$ & \cellcolor[gray]{0.70}$10.53 \pm 4.0\%$ & \cellcolor[gray]{0.76}$10.19 \pm 4.3\%$ & \cellcolor[gray]{0.81}$9.91 \pm 4.9\%$ & \cellcolor[gray]{0.90}$9.42 \pm 4.7\%$ & \cellcolor[gray]{0.90}$9.45 \pm 4.9\%$ & \cellcolor[gray]{0.95}$9.14 \pm 5.8\%$ & \cellcolor[gray]{0.98}$9.00 \pm 5.7\%$ & \cellcolor[gray]{1.00}$8.89 \pm 6.3\%$ \\
    $T_e^\text{edge}$ & \cellcolor[gray]{1.00}$187 \pm 37\%$ & \cellcolor[gray]{0.97}$191 \pm 35\%$ & \cellcolor[gray]{0.92}$201 \pm 30\%$ & \cellcolor[gray]{0.91}$202 \pm 28\%$ & \cellcolor[gray]{0.83}$216 \pm 24\%$ & \cellcolor[gray]{0.78}$225 \pm 21\%$ & \cellcolor[gray]{0.73}$233 \pm 21\%$ & \cellcolor[gray]{0.70}$239 \pm 21\%$ \\
    $T_e^\text{sep}$ & \cellcolor[gray]{1.00}$105 \pm 41\%$ & \cellcolor[gray]{0.99}$106 \pm 37\%$ & \cellcolor[gray]{0.94}$113 \pm 31\%$ & \cellcolor[gray]{0.93}$114 \pm 27\%$ & \cellcolor[gray]{0.85}$125 \pm 24\%$ & \cellcolor[gray]{0.79}$133 \pm 20\%$ & \cellcolor[gray]{0.73}$141 \pm 19\%$ & \cellcolor[gray]{0.70}$145 \pm 18\%$ \\
    $T_e^\text{SOL}$ & \cellcolor[gray]{1.00}$45 \pm 46\%$ & \cellcolor[gray]{0.96}$48 \pm 40\%$ & \cellcolor[gray]{0.92}$51 \pm 34\%$ & \cellcolor[gray]{0.88}$53 \pm 30\%$ & \cellcolor[gray]{0.84}$56 \pm 31\%$ & \cellcolor[gray]{0.81}$58 \pm 29\%$ & \cellcolor[gray]{0.74}$63 \pm 25\%$ & \cellcolor[gray]{0.70}$66 \pm 27\%$ \\
    $T_i^\text{edge}$ & $204 \pm 28\%$ & $205 \pm 28\%$ & $208 \pm 28\%$ & $209 \pm 28\%$ & $209 \pm 29\%$ & $209 \pm 29\%$ & $208 \pm 29\%$ & $206 \pm 30\%$ \\
    $T_i^\text{sep}$ & $162 \pm 29\%$ & $163 \pm 30\%$ & $167 \pm 30\%$ & $168 \pm 30\%$ & $171 \pm 29\%$ & $172 \pm 28\%$ & $173 \pm 27\%$ & $174 \pm 27\%$ \\
    $T_i^\text{SOL}$ & $141 \pm 32\%$ & $144 \pm 31\%$ & $149 \pm 29\%$ & $151 \pm 27\%$ & $155 \pm 26\%$ & $158 \pm 24\%$ & $161 \pm 23\%$ & $167 \pm 23\%$ \\
    $T_i^\text{sep}/T_e^\text{sep}$ & \cellcolor[gray]{0.70}$1.57 \pm 15\%$ & \cellcolor[gray]{0.73}$1.53 \pm 13\%$ & \cellcolor[gray]{0.81}$1.43 \pm 15\%$ & \cellcolor[gray]{0.81}$1.42 \pm 13\%$ & \cellcolor[gray]{0.89}$1.31 \pm 17\%$ & \cellcolor[gray]{0.93}$1.26 \pm 15\%$ & \cellcolor[gray]{0.98}$1.21 \pm 12\%$ & \cellcolor[gray]{1.00}$1.18 \pm 14\%$ \\
    $T_i^\text{edge}/T_i^\text{sep}$ & \cellcolor[gray]{0.70}$1.27 \pm 3.6\%$ & \cellcolor[gray]{0.73}$1.26 \pm 3.3\%$ & \cellcolor[gray]{0.82}$1.24 \pm 3.3\%$ & \cellcolor[gray]{0.87}$1.22 \pm 3.3\%$ & \cellcolor[gray]{0.93}$1.21 \pm 3.3\%$ & \cellcolor[gray]{0.95}$1.20 \pm 3.5\%$ & \cellcolor[gray]{1.00}$1.19 \pm 3.8\%$ & \cellcolor[gray]{1.00}$1.19 \pm 3.7\%$ \\
    $T_i^\text{SOL}/T_i^\text{sep}$ & \cellcolor[gray]{1.00}$0.88 \pm 3.5\%$ & \cellcolor[gray]{0.98}$0.89 \pm 2.4\%$ & \cellcolor[gray]{0.95}$0.89 \pm 3.1\%$ & \cellcolor[gray]{0.92}$0.90 \pm 4.0\%$ & \cellcolor[gray]{0.88}$0.91 \pm 5.5\%$ & \cellcolor[gray]{0.86}$0.91 \pm 5.8\%$ & \cellcolor[gray]{0.77}$0.94 \pm 6.3\%$ & \cellcolor[gray]{0.70}$0.95 \pm 7.5\%$ \\
    $T_{i,\perp}^\text{sep}/T_{i,\parallel}^\text{sep}$ & \cellcolor[gray]{1.00}$1.38 \pm 8.0\%$ & \cellcolor[gray]{0.92}$1.40 \pm 7.7\%$ & \cellcolor[gray]{0.89}$1.41 \pm 7.4\%$ & \cellcolor[gray]{0.72}$1.46 \pm 6.5\%$ & \cellcolor[gray]{0.80}$1.44 \pm 6.8\%$ & \cellcolor[gray]{0.76}$1.45 \pm 5.5\%$ & \cellcolor[gray]{0.70}$1.47 \pm 5.7\%$ & \cellcolor[gray]{0.70}$1.47 \pm 4.3\%$ \\
    $T_{e,\perp}^\text{sep}/T_{e,\parallel}^\text{sep}$ & $1.13 \pm 8.6\%$ & $1.13 \pm 7.7\%$ & $1.13 \pm 6.9\%$ & $1.14 \pm 5.6\%$ & $1.16 \pm 5.2\%$ & $1.16 \pm 5.0\%$ & $1.17 \pm 5.1\%$ & $1.18 \pm 4.8\%$ \\
    \hline
    \multicolumn{9}{l}{\textit{Normalized gradients, Eq.~\eqref{eq:norm_gradient}}} \\
    $R/L_{n_e}$ & \cellcolor[gray]{1.00}$23.7 \pm 2.9\%$ & \cellcolor[gray]{0.92}$24.5 \pm 3.3\%$ & \cellcolor[gray]{0.86}$25.1 \pm 3.6\%$ & \cellcolor[gray]{0.81}$25.6 \pm 2.9\%$ & \cellcolor[gray]{0.80}$25.7 \pm 3.2\%$ & \cellcolor[gray]{0.74}$26.2 \pm 3.1\%$ & \cellcolor[gray]{0.72}$26.5 \pm 2.9\%$ & \cellcolor[gray]{0.70}$26.7 \pm 3.0\%$ \\
    $R/L_{T_e}$ & \cellcolor[gray]{0.70}$16.5 \pm 6.9\%$ & \cellcolor[gray]{0.76}$16.0 \pm 7.4\%$ & \cellcolor[gray]{0.82}$15.6 \pm 8.9\%$ & \cellcolor[gray]{0.83}$15.5 \pm 7.7\%$ & \cellcolor[gray]{0.90}$15.0 \pm 10\%$ & \cellcolor[gray]{0.94}$14.7 \pm 9.0\%$ & \cellcolor[gray]{0.98}$14.4 \pm 7.1\%$ & \cellcolor[gray]{1.00}$14.2 \pm 8.2\%$ \\
    $R/L_{T_i}$ & \cellcolor[gray]{0.70}$7.7 \pm 14\%$ & \cellcolor[gray]{0.73}$7.5 \pm 13\%$ & \cellcolor[gray]{0.82}$7.0 \pm 14\%$ & \cellcolor[gray]{0.87}$6.6 \pm 15\%$ & \cellcolor[gray]{0.92}$6.2 \pm 15\%$ & \cellcolor[gray]{0.95}$6.1 \pm 17\%$ & \cellcolor[gray]{1.00}$5.7 \pm 20\%$ & \cellcolor[gray]{1.00}$5.7 \pm 21\%$ \\
    \hline
    \multicolumn{9}{l}{\textit{Fluctuation levels}} \\
    $\delta n_{e,\text{rms}}/\bar n_e|_\text{edge}$ & \cellcolor[gray]{1.00}$0.145 \pm 7.7\%$ & \cellcolor[gray]{0.84}$0.151 \pm 7.0\%$ & \cellcolor[gray]{0.82}$0.152 \pm 6.7\%$ & \cellcolor[gray]{0.70}$0.156 \pm 6.2\%$ & \cellcolor[gray]{0.74}$0.155 \pm 5.4\%$ & \cellcolor[gray]{0.74}$0.155 \pm 4.7\%$ & \cellcolor[gray]{0.77}$0.154 \pm 6.1\%$ & \cellcolor[gray]{0.70}$0.156 \pm 7.3\%$ \\
    $\delta n_{e,\text{rms}}/\bar n_e|_\text{SOL}$ & $0.242 \pm 13\%$ & $0.249 \pm 12\%$ & $0.240 \pm 13\%$ & $0.257 \pm 11\%$ & $0.248 \pm 12\%$ & $0.245 \pm 11\%$ & $0.254 \pm 12\%$ & $0.259 \pm 9.8\%$ \\
    $\delta T_{e,\text{rms}}/\bar T_e|_\text{edge}$ & \cellcolor[gray]{0.70}$0.050 \pm 15\%$ & \cellcolor[gray]{0.76}$0.047 \pm 13\%$ & \cellcolor[gray]{0.82}$0.044 \pm 11\%$ & \cellcolor[gray]{0.86}$0.042 \pm 11\%$ & \cellcolor[gray]{0.92}$0.039 \pm 10\%$ & \cellcolor[gray]{0.95}$0.037 \pm 8.0\%$ & \cellcolor[gray]{0.97}$0.036 \pm 6.6\%$ & \cellcolor[gray]{1.00}$0.034 \pm 9.1\%$ \\
    $\delta T_{e,\text{rms}}/\bar T_e|_\text{SOL}$ & $0.199 \pm 14\%$ & $0.203 \pm 11\%$ & $0.199 \pm 9.5\%$ & $0.213 \pm 9.0\%$ & $0.207 \pm 9.0\%$ & $0.206 \pm 10\%$ & $0.212 \pm 7.5\%$ & $0.206 \pm 7.5\%$ \\
    $\delta T_{i,\text{rms}}/\bar T_i|_\text{edge}$ & \cellcolor[gray]{0.70}$0.050 \pm 11\%$ & \cellcolor[gray]{0.77}$0.048 \pm 12\%$ & \cellcolor[gray]{0.81}$0.047 \pm 12\%$ & \cellcolor[gray]{0.86}$0.046 \pm 13\%$ & \cellcolor[gray]{0.90}$0.044 \pm 13\%$ & \cellcolor[gray]{0.95}$0.043 \pm 14\%$ & \cellcolor[gray]{0.98}$0.042 \pm 12\%$ & \cellcolor[gray]{1.00}$0.042 \pm 11\%$ \\
    $\delta T_{i,\text{rms}}/\bar T_i|_\text{SOL}$ & \cellcolor[gray]{0.70}$0.103 \pm 25\%$ & \cellcolor[gray]{0.77}$0.097 \pm 26\%$ & \cellcolor[gray]{0.85}$0.090 \pm 26\%$ & \cellcolor[gray]{0.91}$0.085 \pm 25\%$ & \cellcolor[gray]{0.98}$0.079 \pm 24\%$ & \cellcolor[gray]{0.97}$0.080 \pm 19\%$ & \cellcolor[gray]{1.00}$0.078 \pm 16\%$ & \cellcolor[gray]{1.00}$0.078 \pm 15\%$ \\
    $\delta n_{e,\text{rms}}/\bar n_e|_{r/a=0.9}$ & $0.054 \pm 16\%$ & $0.053 \pm 18\%$ & $0.053 \pm 19\%$ & $0.056 \pm 19\%$ & $0.059 \pm 19\%$ & $0.059 \pm 18\%$ & $0.062 \pm 16\%$ & $0.062 \pm 18\%$ \\
    \hline
    \multicolumn{9}{l}{\textit{Separatrix transport}} \\
    $\Gamma_e^\text{sep}$ & $10.59 \pm 13\%$ & $10.35 \pm 20\%$ & $10.26 \pm 15\%$ & $10.36 \pm 14\%$ & $10.80 \pm 12\%$ & $10.18 \pm 12\%$ & $10.50 \pm 14\%$ & $10.56 \pm 14\%$ \\
    $q_e^\text{sep}$ & $22.0 \pm 28\%$ & $22.9 \pm 25\%$ & $22.9 \pm 22\%$ & $22.3 \pm 22\%$ & $23.5 \pm 20\%$ & $24.1 \pm 13\%$ & $25.0 \pm 14\%$ & $24.5 \pm 17\%$ \\
    $q_i^\text{sep}$ & $0.0 \pm 2.8$ & $-0.1 \pm 2.8$ & $-0.6 \pm 2.8$ & $-1.3 \pm 2.3$ & $-0.8 \pm 2.3$ & $-1.2 \pm 2.3$ & $-1.9 \pm 80\%$ & $-2.0 \pm 81\%$ \\
    $D_e$ & \cellcolor[gray]{0.70}$1.22 \pm 16\%$ & \cellcolor[gray]{0.73}$1.20 \pm 25\%$ & \cellcolor[gray]{0.83}$1.14 \pm 20\%$ & \cellcolor[gray]{0.90}$1.10 \pm 19\%$ & \cellcolor[gray]{0.80}$1.16 \pm 16\%$ & \cellcolor[gray]{0.97}$1.05 \pm 14\%$ & \cellcolor[gray]{0.99}$1.03 \pm 17\%$ & \cellcolor[gray]{1.00}$1.03 \pm 17\%$ \\
    $\chi_e$ & $3.70 \pm 20\%$ & $4.07 \pm 23\%$ & $4.15 \pm 22\%$ & $4.18 \pm 22\%$ & $4.45 \pm 19\%$ & $4.31 \pm 18\%$ & $4.63 \pm 18\%$ & $4.73 \pm 22\%$ \\
    $D_i$ & $0.85 \pm 58\%$ & $0.79 \pm 62\%$ & $0.77 \pm 56\%$ & $0.82 \pm 46\%$ & $0.92 \pm 36\%$ & $0.80 \pm 35\%$ & $0.77 \pm 39\%$ & $0.85 \pm 35\%$ \\
    $\chi_i$ & \cellcolor[gray]{0.70}$0.01 \pm 0.95$ & \cellcolor[gray]{0.71}$-0.03 \pm 1.05$ & \cellcolor[gray]{0.76}$-0.23 \pm 1.13$ & \cellcolor[gray]{0.82}$-0.47 \pm 0.98$ & \cellcolor[gray]{0.79}$-0.36 \pm 0.95$ & \cellcolor[gray]{0.86}$-0.66 \pm 0.96$ & \cellcolor[gray]{0.93}$-0.95 \pm 62\%$ & \cellcolor[gray]{1.00}$-1.25 \pm 51\%$ \\
    $q_e/(\tfrac{3}{2}\overline{T}_e\Gamma_e)$ & \cellcolor[gray]{0.70}$0.91 \pm 16\%$ & \cellcolor[gray]{0.77}$0.86 \pm 24\%$ & \cellcolor[gray]{0.80}$0.84 \pm 21\%$ & \cellcolor[gray]{0.83}$0.81 \pm 17\%$ & \cellcolor[gray]{0.91}$0.75 \pm 19\%$ & \cellcolor[gray]{0.92}$0.75 \pm 17\%$ & \cellcolor[gray]{0.95}$0.72 \pm 17\%$ & \cellcolor[gray]{1.00}$0.69 \pm 18\%$ \\
    $\lambda_q$ & $3.94 \pm 31\%$ & $4.10 \pm 26\%$ & $4.28 \pm 25\%$ & $4.00 \pm 32\%$ & $3.90 \pm 22\%$ & $3.89 \pm 25\%$ & $4.01 \pm 17\%$ & $3.60 \pm 28\%$ \\
    \hline
    \multicolumn{9}{l}{\textit{Global quantities}} \\
    $\tau_E$ & \cellcolor[gray]{1.00}$0.76 \pm 15\%$ & \cellcolor[gray]{0.95}$0.80 \pm 16\%$ & \cellcolor[gray]{0.90}$0.83 \pm 19\%$ & \cellcolor[gray]{0.85}$0.86 \pm 19\%$ & \cellcolor[gray]{0.82}$0.89 \pm 22\%$ & \cellcolor[gray]{0.82}$0.89 \pm 18\%$ & \cellcolor[gray]{0.70}$0.97 \pm 16\%$ & \cellcolor[gray]{0.76}$0.93 \pm 21\%$ \\
    $\delta\dot W/P_\text{in}$ & \cellcolor[gray]{0.99}$0.056 \pm 39\%$ & \cellcolor[gray]{1.00}$0.055 \pm 59\%$ & \cellcolor[gray]{0.90}$0.068 \pm 42\%$ & \cellcolor[gray]{0.94}$0.063 \pm 63\%$ & \cellcolor[gray]{0.98}$0.058 \pm 78\%$ & \cellcolor[gray]{0.80}$0.082 \pm 45\%$ & \cellcolor[gray]{0.80}$0.081 \pm 41\%$ & \cellcolor[gray]{0.70}$0.095 \pm 58\%$ \\
    $\Delta t$ & \cellcolor[gray]{0.70}$3.59 \pm 15\%$ & \cellcolor[gray]{0.70}$3.60 \pm 14\%$ & \cellcolor[gray]{0.77}$3.39 \pm 13\%$ & \cellcolor[gray]{0.78}$3.35 \pm 11\%$ & \cellcolor[gray]{0.83}$3.20 \pm 10\%$ & \cellcolor[gray]{0.88}$3.03 \pm 9.3\%$ & \cellcolor[gray]{0.93}$2.88 \pm 9.0\%$ & \cellcolor[gray]{1.00}$2.67 \pm 9.2\%$ \\
    $t_\text{end}$ & \cellcolor[gray]{0.70}$1.99 \pm 8.6\%$ & \cellcolor[gray]{0.72}$1.95 \pm 9.1\%$ & \cellcolor[gray]{0.74}$1.91 \pm 11\%$ & \cellcolor[gray]{0.77}$1.83 \pm 14\%$ & \cellcolor[gray]{0.85}$1.67 \pm 15\%$ & \cellcolor[gray]{0.89}$1.58 \pm 13\%$ & \cellcolor[gray]{0.96}$1.44 \pm 12\%$ & \cellcolor[gray]{1.00}$1.36 \pm 11\%$ \\
    \hline
    \hline\hline
    \end{tabularx}
\end{table*}

\clearpage

\bibliographystyle{apsrev4-2}
\bibliography{references}

\end{document}